\documentclass[10pt, twocolumn]{iopart}

\usepackage{comment}
\usepackage{xcolor}
\usepackage{array}
\usepackage{arydshln}
\expandafter\let\csname equation*\endcsname=\relax 
\expandafter\let\csname endequation*\endcsname=\relax 
\usepackage{amsmath}
\usepackage{amssymb}
\usepackage{subcaption}
\usepackage{gensymb}
\usepackage{cancel}
\usepackage[margin=0.6in]{geometry}
\usepackage{makecell}
\usepackage{graphicx,booktabs}
\usepackage{multicol}
\usepackage{url}

\usepackage[separate-uncertainty=true]{siunitx}
\DeclareSIUnit\torr{torr}

\begin{document}

\twocolumn[
  \begin{@twocolumnfalse}
    \title{Determination of confinement regime boundaries via separatrix parameters on Alcator C-Mod based on a model for interchange-drift-Alfvén turbulence}

\author{M.A. Miller$^1$,
        J.W. Hughes$^1$,
        T. Eich$^2$,
        G.R. Tynan$^3$,
        P. Manz$^4$,
        T. Body$^2$,
        D. Silvagni$^5$,
        O. Grover$^5$,
        A.E. Hubbard$^1$,
        A. Cavallaro$^1$,
        M. Wigram$^1$,
        A.Q. Kuang$^2$,
        S. Mordijck$^6$,
        B. LaBombard$^1$,
        J. Dunsmore$^1$,
        D. Whyte$^1$}

\address{$^1$MIT Plasma Science and Fusion Center, Cambridge, MA 02139, USA
}
\address{$^2$Commonwealth Fusion Systems, Devens, MA 01434, USA
}
\address{$^3$University of California, San Diego, CA 92093, USA
}
\address{$^4$Institute of Physics, University of Greifswald, Felix-Hausdorff-Str.6, Greifswald, 17489, Germany
}
\address{$^5$Max-Planck-Institut für Plasmaphysik, Boltzmannstraße 2, D-85748, Garching, Germany
}
\address{$^6$William \& Mary, Williamsburg, VA 23188, USA
}
\ead{millerma@mit.edu}

\maketitle
    \label{sec:abstract}

\begin{abstract}
The separatrix operational space (SepOS) model [Eich \& Manz, \emph{Nuclear Fusion} (2021)] is shown to predict the L-H transition, the L-mode density limit, and the ideal MHD ballooning limit in terms of separatrix parameters for a wide range of Alcator C-Mod plasmas. The model is tested using Thomson scattering measurements across a wide range of operating conditions on C-Mod, spanning $\overline{n}_{e} = 0.3 - 5.5 \times 10^{20}$m$^{-3}$, $B_{t} = 2.5 - 8.0$ T, and $B_{p} = 0.1 - 1.2$ T. An empirical regression for the electron pressure gradient scale length, $\lambda_{p_{e}}$, against a turbulence control parameter, $\alpha_{t}$, and the poloidal fluid gyroradius, $\rho_{s,p}$, for H-modes is constructed and found to require positive exponents for both regression parameters, indicating turbulence widening of near-SOL widths at high $\alpha_{t}$ and an inverse scaling with $B_{p}$, consistent with results on AUG. The SepOS model is also tested in the unfavorable drift direction and found to apply well to all three boundaries, including the L-H transition as long as a correction to the Reynolds energy transfer term, $\alpha_\mathrm{RS} < 1$ is applied. I-modes typically exist in the unfavorable drift direction for values of $\alpha_{t} \lesssim 0.35$. Finally, an experiment studying the transition between the type-I ELMy and EDA H-mode is analyzed using the same framework. It is found that a recently identified boundary at $\alpha_{t} = 0.55$ excludes most EDA H-modes but that the balance of wavenumbers responsible for the L-mode density limit, namely $k_\mathrm{EM} = k_\mathrm{RBM}$, may better describe the transition on C-Mod. The ensemble of boundaries validated and explored is then applied to project regime access and limit avoidance for the SPARC primary reference discharge parameters.
\end{abstract}

  \end{@twocolumnfalse}
]

\section{Introduction}
\label{sec:intro}

The success of fusion reactors will depend on their ability to access desirable operational regimes and avoid detrimental operational limits. A reactor must be designed to maximize performance but not at the cost of sacrificing power exhaust capabilities and especially not at the cost of plasma disruption. This balancing act is made all the more challenging by an incomplete physics understanding of the fundamental processes responsible for regime transitions and disruptive limits. The transition between the low-confinement mode (L-mode) and the high-confinement mode (H-mode), for example, is associated with high auxiliary power, steepening ion pressure gradients, and large radial electric field shear \cite{burrell_effects_1997}, but in the absence of a fully predictive model for the transition, future devices rely on extrapolation of power threshold scalings. The well-known L-mode density limit (LDL) \cite{greenwald_new_1988, petrie_plasma_1993} similarly lacks a robust physical understanding, and a number of experiments have been observed to reach values higher than the most commonly used metric for the density limit \cite{gibson_fusion_1990, kamada_study_1991, stabler_density_1992, hong_dl_2023, thome_overview_2024}, the Greenwald density \cite{greenwald_new_1988}. Finally, limits to H-mode operation at high density like the so-called H-mode density limit (HDL) \cite{petrie_plasma_1993, greenwald_density_2002} are not fully understood, making avoidance prediction for future devices difficult.

Recently, the separatrix operational space model (SepOS) has been developed to explore boundaries for regime transitions and limits of the tokamak operational space at the separatrix \cite{eich_separatrix_2021}. The SepOS model is built on parameters from interchange-drift-Alvén (DALF) turbulence and draws from the works of Scott \cite{scott_drift_2005} and Rogers, Drake, and Zeiler \cite{rogers_phase_1998}. The model from Eich and Manz well separates regime transitions and limits on ASDEX Upgrade (AUG). The current paper represents the first validation of the SepOS model on a device other than AUG. Using measurements from the edge Thomson Scattering system on Alcator C-Mod, it is shown that that SepOS model also applies well to the operational space of C-Mod.

The SepOS model has proven successful for describing three main boundaries in the AUG operational space at the separatrix: the transition from L-mode to H-mode (L-H transition), the LDL, and the ideal magnetohydrodynamic (MHD) limit. Application of the model involved the observation that the plasma scale lengths were inversely proportional to the poloidal magnetic field, $B_{p}$, and proportional to a collisionality-like turbulence control parameter, $\alpha_{t}$, introduced in \cite{eich_turbulence_2020} and given by $\alpha_{t} = 2.98 \times 10^{-18} R_\mathrm{geo}\hat{q}_\mathrm{cyl}^{2} \frac{n_{e}}{T_{e}^{2}} Z_\mathrm{eff}$ for Alcator C-Mod parameters, where $R_\mathrm{geo}$ is the geometric major radius, assumed here to be equal to the device major radius, $R_{0}$, and $\hat{q}_\mathrm{cyl}$ is the cylindrical safety factor, calculated according to $\hat{q}_\mathrm{cyl} = \frac{B_{t}}{B_{p}} \times \frac{\hat{\kappa}}{R_\mathrm{geo}/a_\mathrm{geo}}$, where $B_{t}$ and $B_{p}$ are the toroidal and poloidal magnetic fields respectively, $a_\mathrm{geo}$ is the geometric minor radius, taken equal to the device minor radius, and $\hat{\kappa}$ is the effective elongation, defined in \cite{eich_separatrix_2021}. Note that $\alpha_{t}$ differs from the often used collisionality, $\nu^{*}$, by a factor of $\hat{q}_\mathrm{cyl}$, as well as the diamagnetic parameter from the work of Rogers, Drake, and Zeiler, $\alpha_{d}$, by a factor of $\frac{\lambda_{n_{e}}}{R_\mathrm{geo}}$. Further analysis then focused on limits to operation at the highest densities in both H-mode and L-mode, and these limits from the SepOS model were compared to those from other models for disruptive high density operation \cite{manz_power_2023}. Recent work on AUG continues analysis of H-modes specifically, concluding that the transition between type of H-mode can also be described by movement across the SepOS, in particular as $\alpha_{t}$ varies \cite{faitsch_analysis_2023}. Another study has extended the model to discharges in the unfavorable drift direction on AUG, introducing a correction factor to the model for the L-H criterion when the $\nabla B$-drift points in the unfavorable direction \cite{grover_reduced_2024}. 

This paper finds similar phenomenology in the C-Mod SepOS. Using the same DALF-normalized dimensionless quantities, the L-H transition, the LDL, and the ideal MHD limits are all well-described by the SepOS model. As on AUG, it is observed that the plasma scale lengths near the separatrix widen with $\alpha_{t}$ in H-mode (consistent with earlier work that found the same to be the case just outside the separatrix in L-mode \cite{labombard_evidence_2005}). A scaling for this gradient scale length widening is then used to translate the boundaries in dimensionless space into the more intuitive dimensional space in terms of electron density, $n_{e}$, and electron temperature, $T_{e}$. Further work probes extension of the model to include C-Mod discharges in the unfavorable $\nabla B$-drift direction as well as to elucidate the transition between the H-mode type.

Below, we recall earlier work in edge turbulence theory application to operational boundaries in Section \ref{sec:history}, and describe how the SepOS model improves on earlier attempts at prediction. Section \ref{sec:thomson} initiates analysis by describing the methodology for evaluating separatrix quantities in their gradient scale lengths, introducing also the scaling for the widening of gradient scale lengths with $\alpha_{t}$. From these, Section \ref{sec:forward_field_os} shows computation of SepOS quantities and how these apply to a large database on Alcator C-Mod. Section \ref{sec:reverse_field_os} does the same for discharges in the unfavorable drift direction, highlighting differences in H-mode and I-mode access in this configuration. Section \ref{sec:elmy_eda} explores the H-mode space further, describing analysis in the transition between H-mode type. Finally, Section \ref{sec:sparc_projections} describes model application to SPARC and Section \ref{sec:conclusions} concludes and remarks on next steps for model development.


\section{A reduced turbulence model for L- and H-mode existence}
\label{sec:history}

\subsection{Early theories for fluid edge turbulence and examination on Alcator C-Mod}

The development of a description of the tokamak edge using electromagnetic fluid drift turbulence (EMFDT) theory began with the works of Scott and Rogers, Drake, and Zeiler (RDZ). Scott began studying the nonlinear drift wave instability through the development of an equation set describing DALF turbulence, \cite{scott_three-dimensional_1997, scott_nonlinear_2002, scott_drift_2005}, which capture the effects of both interchange turbulence and MHD instabilities. From these equations, he introduced a number of dimensionless parameters, including a normalized pressure gradient, $\hat{\beta}$, and a collisionality parameter, $C_{0}$. Simultaneously, Rogers, Drake, and Zeiler carried out 3D simulations of the Braginskii equations and identified two parameters strongly influential in determining edge turbulence \cite{rogers_enhancement_1997, rogers_phase_1998}. These were the ballooning parameter, $\alpha$ (or $\alpha_\mathrm{MHD}$), and the diamagnetic parameter, $\alpha_\mathrm{d}$. Similarly to the work of Scott, the former parameter is a normalized pressure gradient and the second parameter has strong dependence on collisionality. The RDZ work suggested the edge plasma phase space could be described by these two parameters and that both the density limit and the transition to H-mode occurred at high values of $\alpha$. The former also required low $\alpha_{d}$ and the latter high $\alpha_{d}$.

The current paper adds to a number of validation studies of the models discussed above. These tests have been conducted on a number of machines, including COMPASS-D, JET, NSTX, DIII-D, and AUG \cite{connor_review_2000, righi_comparison_2000, kaye_low-_2003, guzdar_low_2004, eich_turbulence_2020}. The models have also been tested using Alcator C-Mod data. Initial attempts on C-Mod to provide experimental verification of the theories of Scott and RDZ involved the use of electron cyclotron emission measurements of the pedestal for calculation of $\alpha$ and $\alpha_{d}$ \cite{hubbard_local_1998}. As in the RDZ theory, that work found good separation between L- and H-modes using those parameters. Later work involved calculation of these same parameters but with Thomson scattering (TS) measurements. This was done first at mid-pedestal \cite{hughes_edge_2007} and then at both the top of the pedestal and at the separatrix \cite{ma_scaling_2012}, turning attention also to profile analysis near the L-H transition specifically.

One of the most extensive comparisons with EMFDT theory from both Scott and RDZ involved the use of scanning Langmuir probes to study the phase-space of the near-SOL \cite{labombard_evidence_2005}. This work computed key control parameters from both the RDZ and Scott theories and examined their dependence, showing consistency between the RDZ parameters mentioned above and similar parameters from Scott, in particular $\hat{\beta}$ and $C_{0}$. Using this combination of parameters, the work identified a clear boundary that separated ohmic L-modes from ohmic H-modes, corresponding to the L-H transition, and an inaccessible region at high $\alpha_\mathrm{MHD}$ and low $\alpha_{d}$, corresponding to operation limited at high density in both L-mode and H-mode. The work also found that linear dependence on the safety factor, $q$ and square root dependence on the electron-ion collision mean free path, $\lambda_\mathrm{ei}$, normalized to the major radius $R$, best organized near-SOL pressure gradient scale lengths, across a range of $q_{95}$. Later work codified this dependence on collisionality into the $\Lambda$ parameter \cite{labombard_critical_2008}, similar to $\alpha_{t}$, and found a great deal of success in its use for defining operational boundaries.

\subsection{The SepOS model}
\label{sec:SepOS}

The SepOS model from Eich and Manz \cite{eich_separatrix_2021} begins from the DALF equation set proposed by Scott. These are detailed in the works of Scott referenced above and are summarized in the appendix of \cite{eich_separatrix_2021}. The equations are then normalized according to characteristic time scales, length scales, and speeds. From this normalization, wavenumbers and energy transfer rates (or equivalently growth rates as suggested in \cite{grover_reduced_2024}) of particular importance are extracted and compared. The balance of these quantities of interest underpin changes in turbulence and become the ingredients for the boundaries in the SepOS model.

In contrast with the previous models that do not specify a particular location of importance in the plasma, the SepOS model proposes the region near the separatrix as a key location for determining the edge plasma turbulence properties, which define the plasma operational regime. Indeed, all SepOS boundaries are derived in terms of dimensionless quantities at this radial location. Applicability of the model to tokamak operation, however, benefits from the translation of dimensionless quantities into the more familiar ($n_{e}, T_{e}$) space. For most of the normalized quantities in SepOS, this mapping to dimensional variables is fairly straightforward. A notable exception, however, is the perpendicular length scale, $\lambda_{\perp}$, for which a direct parametrization in terms of $n_{e}$, $T_{e}$, and global plasma parameters is unresolved. This obstacle was circumvented with the use of the scaling developed in \cite{eich_turbulence_2020}. It was observed that plasma scale lengths near the separatrix depend strongly on $\alpha_{t}$ (itself a function of $n_{e}$ and $T_{e}$), as well as the fluid gyroradius, $\rho_{s,p}$. Constructing a scaling for $\lambda_{\perp} = \lambda_{p_{e}}(\rho_{s,p}, \alpha_{t})$ provides the necessary mapping to translate the dimensionless SepOS model into a dimensional guide for directing tokamak plasmas towards a particular regime or away from a particular limit.

\section{Experimental procedure for evaluating separatrix parameters and plasma gradients}
\label{sec:thomson}

\begin{figure*}[h]
\centering
\includegraphics[width=1.6\columnwidth]{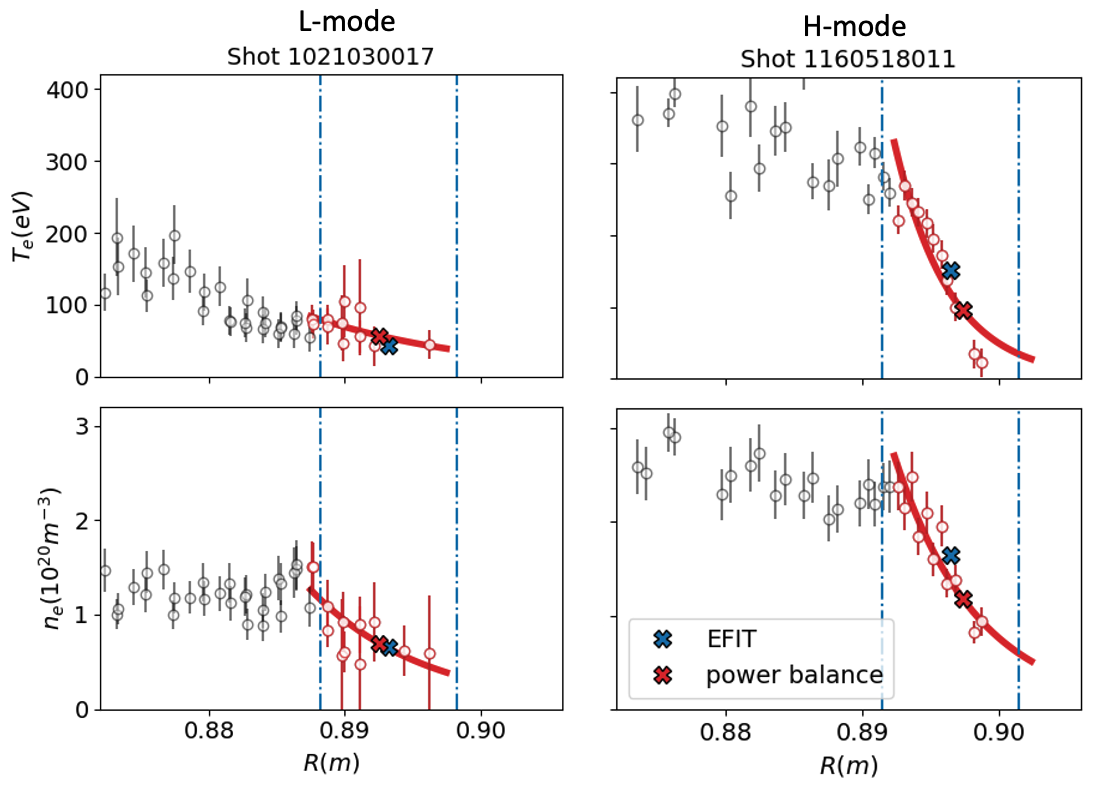}
\caption{Typical profiles measured by edge Thomson Scattering on C-Mod, for both an L-mode (left) and an H-mode (right). Open circles show raw measurements of $T_{e}$ (top) and $n_{e}$ (bottom) from the ETS system. Points in red have been used in the final separatrix fit, shown by the solid red curve, as outlined in section \ref{sec:sep_id}. Points in black are unused. Dash-dotted vertical blue lines represent the 10 mm initial fit interval about the location of the separatrix as identified by the EFIT reconstruction. The separatrix position calculated by EFIT is shown as a blue X and the resulting separatrix position found with power balance is given by the red X.}
\label{fig:thomson_profs}
\end{figure*}

In order to test the SepOS model as well as determine a scaling for $\lambda_{\perp} = \lambda_{p_{e}}$ in terms of local separatrix quantities, a procedure similar to that performed on AUG is carried out on Alcator C-Mod. Using only measurements from its edge Thomson scattering (ETS) system, an iterative procedure yields absolute separatrix electron parameters as well as their gradient scale lengths. From these measurements, a scaling for the H-mode electron pressure gradient scale length in terms of both $\rho_{s,p}$ and $\alpha_{t}$ is derived for Alcator C-Mod. The scaling is compared to that found for AUG H-modes \cite{eich_turbulence_2020} and is used in the following section to translate the normalized SepOS boundaries into the dimensional ($n_{e}, T_{e}$) space.

\subsection{Separatrix identification}
\label{sec:sep_id}
The workhorse diagnostic for validating the SepOS model on C-Mod data is the ETS system. This diagnostic was routinely used to diagnose the edge plasma on C-Mod, in particular the pedestal in H-mode. ETS took measurements of both $n_{e}$ and $T_{e}$. Measurements were taken in the upper chamber near the crown of lower single null (LSN) plasmas. They spanned a region of about 3 cm and could diagnose edge profiles with order millimeter resolution when mapped to the midplane \cite{hughes_high-resolution_2001}. While the ETS system was primarily developed to study pedestal physics, this study has found that it did a good job of also resolving the gradient scale lengths across the separatrix. Figure \ref{fig:thomson_profs} shows an example of $T_{e}$ and $n_{e}$ measurements from ETS for both an H-mode and an L-mode, from shots taken 14 years apart.

The location of the separatrix is identified from power balance in the scrape-off layer (SOL) using the ``two-point model" \cite{stangeby_pc_plasma_2000}. To do so, two expressions for the conducted parallel heat flux, $q_\parallel$, are balanced against each other. The first uses an estimate of the power conducted by electrons into the SOL and is given by:

\begin{equation}
    q_{e,\parallel,\mathrm{cond}} = \frac{f_{e,\mathrm{cond}} P_\mathrm{SOL}}{2\pi R \lambda_{q} \frac{B_{p}}{B}}
    \label{eq:qpar_exp}
\end{equation}

where $f_{e,\mathrm{cond}}$ represents the fraction of the power conducted by electrons into the SOL and is set to 0.325, $P_\mathrm{SOL}$ is the net power entering the SOL, calculated using $P_\mathrm{SOL} = P_\mathrm{oh} + \eta_\mathrm{ICRF} P_\mathrm{ICRF} - P_\mathrm{rad}$, where $P_\mathrm{ICRF}$ is the total injected ion cyclotron range of frequencies (ICRF) power and $\eta_\mathrm{ICRF}$ is its efficiency, set to 1 for simplicity across the datasets analyzed. $P_\mathrm{oh}$ is the ohmic power and $P_\mathrm{rad}$ is the power radiated in the core. $R$ is the plasma major radius and $\lambda_{q}$ is the width of the heat flux channel, estimated at the outer midplane from the $T_{e}$ profile as explained in detail below. 

The second expression is that of the conducted parallel heat flux assuming Spitzer-Härm parallel transport to the divertor, given in terms of the parallel temperature gradient. Assuming $T_{e}$ at the divertor is negligible compared to $T_{e}$ upstream ($T_{e,\mathrm{div}}^{7/2} \ll T_{e,\mathrm{up}}^{7/2}$) and that the power enters the SOL uniformly poloidally, this is expressed by:

\begin{equation}
    q_{e,\parallel,\mathrm{SH}} = \frac{4}{7} \kappa_{0,e} \frac{{T_{e,\mathrm{up}}^{7/2}}}{L_{\parallel}}
    \label{eq:qpar_SH}
\end{equation}

where  $\kappa_{0,e}$ is the parallel electron conductivity coefficient and $L_\parallel$ is the connection length to the outer target, estimated using $L_{\parallel} = q_{95} \pi R$, with $q_{95}$, the safety factor at the 95\% flux surface. 

Equating expressions \ref{eq:qpar_exp} and \ref{eq:qpar_SH} and solving for $T_{e,\mathrm{up}}$ gives:

\begin{equation}
    T_{e,\mathrm{up}} = \left(\frac{7 f_{e,\mathrm{cond}} P_\mathrm{SOL} L_{\parallel} }{8\pi R \lambda_{q} \frac{B_{p}}{B} \kappa_{0,e}} \right)^{2/7}
    \label{eq:power_balance}
\end{equation}

the upstream temperature at the separatrix in terms of experimental parameters. For the remainder of the work, this parameter will be referred to instead as $T_{e}^\mathrm{sep}$. The upstream separatrix position is thus identified as the location at which $T_{e}(R_{\mathrm{sep}}) = T_{e}^\mathrm{sep}$. This location is then used to find $n_{e}^\mathrm{sep} = n_{e}(R_{\mathrm{sep}})$, enabled by the simultaneous measurement of $T_e$ and $n_e$ by ETS.

To find $R_\mathrm{sep}$ (and hence $n_{e}^\mathrm{sep}$) from $T_{e}^\mathrm{sep}$, a fit to the ETS data is needed. Following from \cite{eich_correlation_2018, eich_turbulence_2020}, an exponential decay fit, $Y(R) = Y_{0}\mathrm{exp}[\frac{-(R - R_{\mathrm{sep}})}{\lambda_{Y}}]$, centered at $R_\mathrm{sep}$, is chosen to capture the gradients in the plasma profiles at the separatrix. Here, $\lambda_{Y}$ is a fit coefficient directly output by the fitting algorithm. As such, this fit function allows for direct estimation of electron gradient scale lengths under the assumption that the gradient scale length is constant across the separatrix. With the fit to $n_{e}$ and $T_{e}$ in hand, and because the ETS data extends through the near-SOL, the Spitzer-Härm equality relating $\lambda_{q}$ and the $T_{e}$ gradient scale length, $\lambda_{q} = \frac{2}{7}\lambda_{T_{e}}$, is used. This avoids the need to choose a scaling for $\lambda_{q}$ and allows for separatrix identification directly from the profile information itself. Separatrix identification begins by fitting $T_{e}$ with the exponential function introduced above over an interval of 10 mm centered about the separatrix position as estimated by EFIT. Using $\lambda_{T_{e}}$ from the fit, $\lambda_{q}$ is calculated to solve Equation \ref{eq:power_balance}, and a value for $T_{e}^\mathrm{sep}$ (as well as $R_\mathrm{sep}$ and $n_{e}^\mathrm{sep}$) is computed. A new fit is then computed, this time over an interval about $R_\mathrm{sep}$. The new value of $T_{e}^\mathrm{sep}$ is compared to the previous, and the iteration continues until the difference in $T_{e}^\mathrm{sep}$ from one iteration to the next, $\Delta T \leq$ 2 eV.


The procedure described above makes a number of simplifying assumptions, useful for database studies, but which merit further consideration. The first of these is that $T_{e,\mathrm{div}}^{7/2} \ll T_{e,\mathrm{up}}^{7/2}$, which simplifies the equation set, making the solution more numerically tractable as it can be solved directly. Without routinely available target measurements of $T_{e}$ and $n_{e}$, it is difficult to know how strong of an approximation this is. A second potentially impactful approximation is the assumption that these plasmas are in the Spitzer-Härm conductivity regime. While this is valid for highly collisional plasmas typical of C-Mod, the inclusion of kinetic effects may be important for discharges at high power and low density \cite{power_scaling_2023}. Finally, while not an assumption, some uncertainty exists in the value of $f_{e,\mathrm{cond}} = f_{e}f_\mathrm{cond}$, which depends on energy partition between various transport channels. Uncertainty is also present in $P_\mathrm{SOL}$ as a result of uncertainty in $\eta_\mathrm{ICRF}$ and in $P_\mathrm{rad}$. More details on the sources and accounting of these uncertainties can be found in \cite{silvagni_crossmachine_2024}.

\begin{table*}[h]
\begin{center}
\caption{Parameter ranges for datasets used in analysis in Sections \ref{sec:thomson} -- \ref{sec:elmy_eda}.}
\label{tab:datasets}
\begin{tabular}{cccc}
Parameter & Full, favorable & Full, unfavorable & ELMy-EDA \\
& (Section \ref{sec:thomson} -- \ref{sec:forward_field_os}) & (Section \ref{sec:reverse_field_os}) & (Section \ref{sec:elmy_eda}) \\
\midrule
\midrule
$B_{t}$ (T) & $3.0 - 7.8$ & $4.3 - 8.0$ & $5.5 - 5.6$ \\
\midrule
$I_{P}$ (MA) & $0.4 - 1.4$ & $0.5 - 1.7$ & $0.9$ \\
\midrule
$B_{p}$ (T) & $0.3 - 1.0$ & $0.4 - 1.2$ & $0.7$ \\
\midrule
$\overline{n}_{e}$ (10$^{20}$m$^{-3}$) & $0.7 - 4.8$ & $0.5 - 2.9$ & $1.2 - 3.2$ \\
\midrule
$P_\mathrm{SOL}$ (MW) & $0.2 - 5.5$ & $0.6 - 5.5$ & $0.8 - 3.3$ \\
\midrule
$\hat{q}_\mathrm{cyl}$ & $2.5 - 10.8$ & $2.9 - 7.9$ & $4.0 - 4.2$\\
\midrule
$f_\mathrm{GW}$ & $0.1 - 0.6$ & $0.1 - 0.6$ & $0.2 - 0.5$\\
\midrule
\end{tabular}
\end{center}
\end{table*}

To account for this uncertainty, equation \ref{eq:power_balance} is solved using an upper and lower estimate of $P_\mathrm{SOL}$, $P_\mathrm{SOL}^{+/-} = (1 \pm 0.2)[P_\mathrm{oh} + \eta_\mathrm{ICRF} P_\mathrm{ICRF} - (1 \mp 0.2)P_\mathrm{rad}]$ to provide an upper and lower estimate for $T_{e}^\mathrm{sep}$, $T_{e,\mathrm{sep}}^{+/-}$. The maximum value of the difference between each of $T_{e,\mathrm{sep}}^{+/-}$ and $T_{e}^\mathrm{sep}$ is used to characterize its uncertainty, $\Delta T_{e}$. Uncertainty in $n_{e}^\mathrm{sep}$, $\Delta n_{e}$ follows the same approach. Uncertainty in $\lambda_{p_{e}}$, $\Delta \lambda_{p_{e}}$ is estimated directly from the diagonal of the covariance matrix of the fits yielding $\lambda_{n_{e}}$ and $\lambda_{T_{e}}$. Assuming uncorrelated error propagation, $\Delta T_{e}$, $\Delta n_{e}$, and $\Delta \lambda_{p_{e}}$ are all then used to calculate uncertainties in all DALF-normalized quantities discussed in the remainder of this work.



\subsection{Description of datasets}
\label{sec:datasets}

With a scheme for estimating separatrix quantities in hand, a number of datasets with shots containing good ETS measurements are assembled. These datasets span a large range in engineering and global parameters, as well as years of operation. The datasets contains discharges with $B \times \nabla B$ directed both toward and away from the active x-point, cases which we refer to as ``favorable'' and ``unfavorable'' $\nabla B$-drift direction. Ranges in parameters for these datasets are summarized in Table \ref{tab:datasets}. The first dataset, shown in the second column of this table, contains many discharges in the favorable $\nabla B$-drift direction and is used in the remainder of Section \ref{sec:thomson} as well as in Section \ref{sec:forward_field_os}. It contains only L-modes and enahnced $D_\alpha$ (EDA) H-modes. The second dataset contains discharges in the unfavorable drift direction. It is used in the analysis in Section \ref{sec:reverse_field_os} and contains L-modes, some steady-state EDA H-modes, and I-modes. Also present in the unfavorable-drift dataset are some transient H-modes without pedestal regulation via continuous transport or edge-localized modes (ELMs), i.e. ``ELM-free'' H-modes. An overview of these confinement modes on C-Mod may be found in \cite{greenwald_cmod_2014}. Data from these first two datasets have been shown in a number of publications from Alcator C-Mod \cite{hughes_observations_2002, hubbard_h-mode_2007, hubbard_physics_2017, brunner_high-resolution_2018, ballinger_dependence_2022, cavallaro_multimachine_2023, miller_collisionality_2024}, although now the ETS data from the consolidated datasets are analyzed in the same way, according to the procedure outlined in section \ref{sec:sep_id}. The data span 14 years of C-Mod operation. A final dataset, which includes only one run day \cite{diallo_observation_2014, diallo_quasi-coherent_2015}, exhibits a mix of EDA H-modes and conventional H-mode with ELMs and is discussed in detail in Section \ref{sec:elmy_eda}. The equilibrium shapes between the former and the latter data set differ as illustrated in Figure 1 of \cite{hughes_pedestal_2013}. 

\subsection{Turbulence widening of separatrix scale lengths}

\begin{figure*}[h]
\centering
\includegraphics[width=1.6\columnwidth]{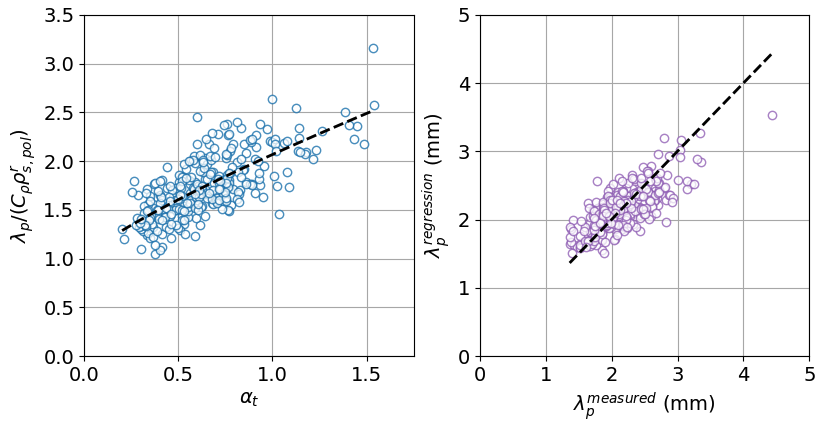}
\caption{Scaling of $\lambda_{p}$ normalized to its dependence on $\rho_{s,p}$ against $\alpha_{t}$ (left) and result of joint $\rho_{s,p}$ and $\alpha_{t}$ regression plotted against experimentally measured value (right). The coefficients used in this scaling are shown in the first column of table \ref{tab:fit_coefficients}.}
\label{fig:lambdap_widening}
\end{figure*}

One of the parameters most important for power handling, setting heat flux spreading on the divertor targets, is the gradient scale length of the parallel heat flux, $q_{\parallel}$, at the separatrix, $\lambda_{q}$ \cite{goldston_heuristic_2012, eich_scaling_2013, chang_gyrokinetic_2017, chen_edge_2017, silvagni_scrape-off_2020, ballinger_dependence_2022}. In the Spitzer-Härm regime, this width is directly linked to the electron temperature gradient scale length, $\lambda_{T_{e}}$ \cite{stangeby_pc_plasma_2000}. A multi-machine study of H-modes found the poloidal magnetic field, $B_{p}$, to organize divertor $\lambda_{q}$ measurements across many devices and a large range in $B_{p}$ \cite{eich_scaling_2013}. A later study solely on AUG \cite{eich_turbulence_2020} found that certain discharges departed from the expectations from the multi-machine scaling, with upstream gradient plasma scale lengths widened up to a factor of three. At low values of $\alpha_{t}$, plasma lengths scaled with $\rho_{s,p} = \frac{\sqrt{m_{i}T_{e}}}{B_{p}}$, as predicted by the multi-machine scaling. As $\alpha_{t} \rightarrow$ 1, the plasma lengths (across channels), increased. It is thought that the widening with $\alpha_{t}$ results from a transition in turbulence type, when interchange effects play a stronger role in drift-wave turbulence \cite{eich_turbulence_2020, eich_separatrix_2021}. In other words, $\alpha_{t}$ mediates the electron adiabaticity, which is responsible for the dephasing of plasma potential fluctuations, $\tilde{\phi}$, and plasma fluctuating quantities, $\tilde{n}$ and $\tilde{T}$ (or $\tilde{p}$ more generally), ultimately increasing plasma transport \cite{scott_nonlinear_2002, scott_drift_2005}.

The technique outlined in the previous section returns $\lambda_{T_{e}}$, as well as electron density gradient scale length, $\lambda_{n_{e}}$. The electron pressure gradient scale length, $\lambda_{p_{e}}$, is then computed using $\lambda_{p_{e}}^{-1} = \lambda_{n_{e}}^{-1} + \lambda_{T_{e}}^{-1}$, from the simple relation for electron pressure, $p_{e} = n_{e}T_{e}$. It is also possible to simply fit $p_{e}$ from ETS directly, and estimate $\lambda_{p_{e}}$ from its fit. The analysis that follows is largely insensitive to choice of $\lambda_{p_{e}}$, but taking the latter approach would increase uncertainty as a result of larger scatter in the measurement. The values from the final iteration of the scheme described in section \ref{sec:sep_id} are used in the remainder of the analysis. For the H-modes in the large, favorable drift direction dataset (second column of Table \ref{tab:datasets}), a multi-variable non-linear least squares regression is performed for $\lambda_{p_{e}}$, using both $\alpha_{t}$ and $\rho_{s,p}$ as the regression variables. A fit is performed for all H-modes in the dataset, using the same proposed regression function as in the AUG work:

\begin{equation}
    \lambda_{p} = (1 + C_{\alpha}\alpha_{t}^{a})C_{\rho}\rho_{s,p}^{r}
    \label{eq:lambdap_widening}
\end{equation}

where $C_{\alpha}$ and $C_{\rho}$ are the regression coefficients for the regressed variables, and $a$ and $r$ are their respective exponents. This fit form captures the observation that at low $\alpha_{t}$, low adiabaticity, the pressure gradient scale lengths are set by the magnetic drifts ($\lambda_{p} \propto \rho_{s,p}$), but that at high adiabaticity, collisional turbulent transport dominates.

Figure \ref{fig:lambdap_widening} shows the regression results. The left panel shows the  measured $\lambda_{p}$, normalized by the $\rho_{s,p}$ dependence, plotted against $\alpha_{t}$. The plot on the right demonstrates the quality of the regression, plotting the regressed against the measured variable. Table \ref{tab:fit_coefficients} compares the regression results for C-Mod H-modes to the regression results from AUG. Compared to AUG, this dataset finds weaker dependence on $\rho_{s,p}$ and $\alpha_{t}$. Possible explanations for the discrepancy are outlined in Section \ref{sec:sparc_projections}. Similar broadening is found for $\lambda_{n}$ and $\lambda_{T}$ independently. Their fit coefficients are included in Table \ref{tab:fit_coefficients} as well, and discrepancy with AUG values is similar to that in $\lambda_{p}$.

\begin{table}[h]
\begin{center}
\caption{Fit coefficients resulting from fitting $\lambda_{p}$, $\lambda_{n}$, and $\lambda_{T}$ against the RHS of equation \ref{eq:lambdap_widening}. Shown also is the comparison for $\lambda_{p}$ on AUG.}
\label{tab:fit_coefficients}
\begin{tabular}{ccccc}
\toprule
Coefficient & \multicolumn{2}{c}{$\lambda_{p}$} & $\lambda_{T}$ & $\lambda_{n}$ \\
\midrule
\midrule
& \underline{C-Mod} & \underline{AUG} & \underline{C-Mod} & \underline{C-Mod} \\
$C_{\alpha}$ & 1.1 & 3.9 & 0.68 & 1.5 \\
$a$ & 0.82 & 1.9 & 0.88 & 0.83 \\
$C_{\rho}$ & 3.6 $\times 10^{-3}$ & 1.3 & 1.6 $\times 10^{-3}$ & 7.0 $\times 10^{-2}$ \\
$r$ & 0.18 & 0.9 & -0.088 & 0.56 \\
\bottomrule
\end{tabular}
\end{center}
\end{table}

Earlier work studying only ohmic L-modes on C-Mod saw a similar effect \cite{labombard_evidence_2005}. It looked at changes to the plasma pressure gradient scale length, measured with Langmuir probes as a function of plasma collision frequency and with different scalings of the safety factor, $q$. Across three values of $q$, this study found that a square dependence, i.e. $q^{2}$, which is the same dependence of $\alpha_{t}$ on $q$, i.e. $\alpha_{t} \propto q^{2}$, best organized the data. A more recent study of C-Mod pedestals finds that the collisionality at the separatrix, $\nu^{*}_\mathrm{sep}$, (or equivalently $\alpha_{t}$, since that study was carried out for only one value of $q$) was responsible for increased pedestal transport \cite{miller_collisionality_2024}. The current study confirms that turbulent broadening of SOL widths at high $\alpha_{t}$ may not only be present in L-modes, but also H-modes, across a large dataset.

\section{The separatrix operational space of Alcator C-Mod in the favorable drift direction}
\label{sec:forward_field_os}

Using the separatrix parameters from the discharges in the same large, favorable drift direction dataset described in Section \ref{sec:datasets}, the three primary boundaries from the SepOS model are tested: the L-H transition, the LDL, and the ideal ballooning MHD limit (IBML). Datasets are sorted by drift direction because while there is no evidence that the LDL and IBML depend on $\nabla B$-drift direction, there is for the L-H transition. Indeed, across a range of devices, it is observed that access to H-mode is considerably facilitated when the ion $\nabla B$-drift points towards the active X-point \cite{ryter_h-mode_1995, hubbard_local_1998, carlstrom_comparison_1998}. This is discussed in greater detail in Section \ref{sec:reverse_field_os}. For this dataset, DALF-normalized wavenumbers and growth rates are calculated. They are described here briefly and in full detail in the following references \cite{eich_turbulence_2020, eich_separatrix_2021, manz_power_2023, faitsch_analysis_2023, grover_reduced_2024}.

\begin{figure}[h]
\centering
\includegraphics[width=\columnwidth]{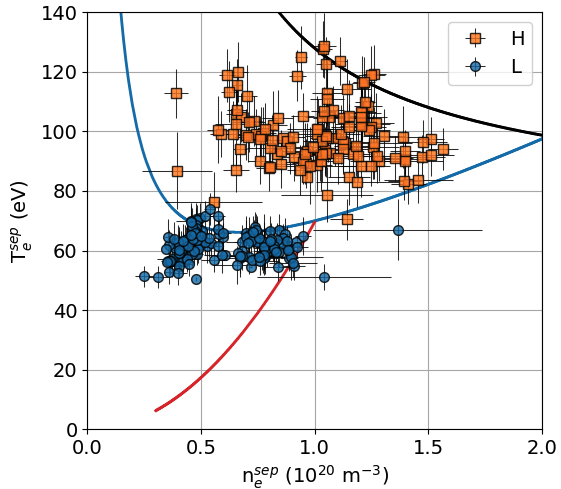}
\caption{Separatrix operational space in terms of $n_{e}$ and $T_{e}$ for typical magnetic equilibrium parameters as listed in Table \ref{tab:mag_geo}. H-modes are shown as orange squares and L-modes are shown as blue circles. Shown also are the three curves in the SepOS model, the L-H curve (blue), the LDL curve (red), and the IBML curve (black). The L-H and IBML curves use the scaling for $\lambda_{p}$ from the first column in Table \ref{tab:fit_coefficients}. The LDL curve uses a fixed $\lambda_{p_{e}}^\mathrm{LDL}$ = 10 mm, the empirically observed value for the highest density L-modes in this range of magnetic equilibrium parameters.}
\label{fig:ff_sepos}
\end{figure}

The criteria for the three boundaries mentioned at the beginning of this section represent a balance of terms -- of energy transfer for the L-H transition and of wavenumbers for the LDL and IBML. Each of these terms can be expressed in terms of easily and routinely measured plasma parameters. These are primarily parameters related to the plasma magnetic equilibrium, including $I_{P}$, the plasma current, $B_{t}$, device size, and the plasma shape, in addition to local plasma parameters at the separatrix, the calculation of which was outlined in Section \ref{sec:thomson}. Holding magnetic equilibrium parameters constant, especially $I_{P}$ and $B_{t}$, allows one to parameterize the criteria solely in terms of $n_{e}$, $T_{e}$, and the selected equilibrium parameters. This then allows for boundary identification in terms of $n_{e}$ and $T_{e}$. In other words, each criterion can be expressed as $f(n_{e}, T_{e}, \mathcal{M}) = 1$, where $f$ is the function for an individual criterion and $\mathcal{M}$ are the selected ``frozen-in" magnetic equilibrium parameters. ($n_{e}$, $T_{e}$) pairs that satisfy $f = 1$ then define the boundaries in ($n_{e}$, $T_{e}$) space. While this expression of the model limits one to a particular choice of $\mathcal{M}$, ``dimensionalizing" the criteria in this way allows one to compare all SepOS boundaries on a single plot, and also helps to build intuition for what control room actuators may drive a plasma towards a particular regime or limit. Figure \ref{fig:ff_sepos} shows $n_{e}^\mathrm{sep}$ and $T_{e}^\mathrm{sep}$ for a subset of plasmas from the first dataset outlined in Section \ref{sec:datasets} at fixed $\mathcal{M}$, as summarized in Table \ref{tab:mag_geo}, as well as the ``dimensionalized" SepOS boundaries, explained in greater detail in the remaining sections. L-modes are shown as blue circles. H-modes are shown as orange squares. The L-H curve is shown in blue. The LDL is shown in red. The IBML is shown in black.

\begin{table}[h]
\begin{center}
\caption{Range in magnetic equilibrium parameters for data shown in Figure \ref{fig:ff_sepos} (and later Figure \ref{fig:rf_sepos}).}
\label{tab:mag_geo}
\begin{tabular}{cc}
Parameter & Value \\
\midrule
\midrule
$B_{t}$ (T) & $5.32 - 5.49$ \\
\midrule
$I_{P}$ (MA) & $0.76 - 0.85$ \\
\midrule
$\hat{q}_\mathrm{cyl}$ & $4.80 - 6.35$ \\
\midrule
$\kappa$ & $1.5 - 1.77$ \\
\midrule
$\delta$ & $0.34 - 0.57$ \\
\midrule
\end{tabular}
\end{center}
\end{table}

\begin{figure*}[h]
\centering
\includegraphics[width=2.0\columnwidth]{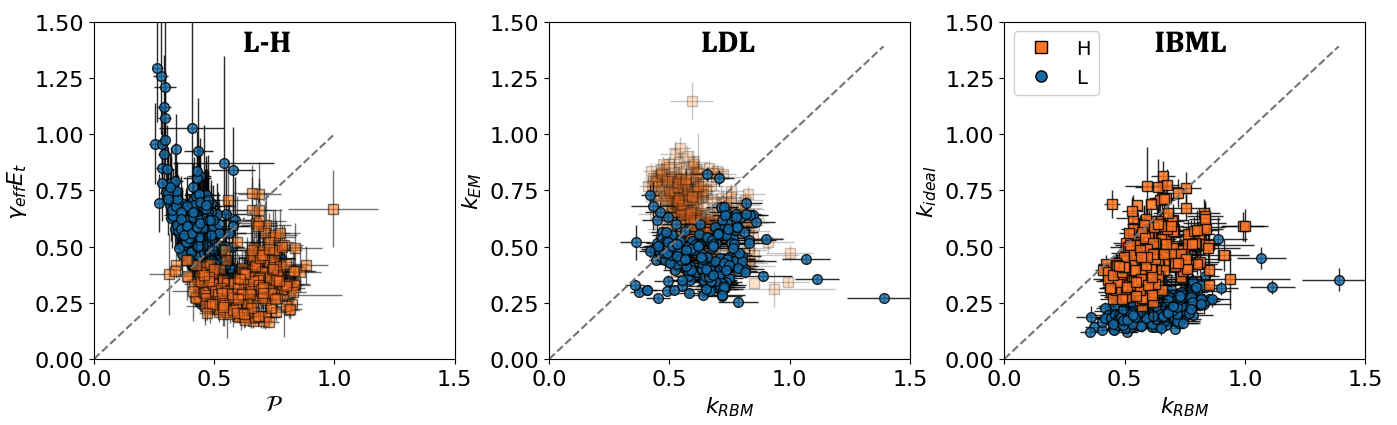}
\caption{The three primary boundaries in the SepOS model in dimensionless terms. Data are from the dataset introduced in section \ref{sec:datasets} and span the ranges in parameters listed in table \ref{tab:datasets}. From left to right are the L-H, LDL, and IBML boundaries, given in dimensionless form by equations \ref{eq:lh_specific}, \ref{eq:LDL}, and \ref{eq:IBML}, respectively. Orange squares represent H-modes and blue circles represent L-modes. The L-H and IBML apply to both L- and H-modes. The LDL only applies to L-modes. Equality lines are shown as dashed gray lines.}
\label{fig:sepos_norm}
\end{figure*}

\subsection{L-H transition}
\label{sec:lh}

The SepOS model posits that an H-mode can be sustained if the transfer of energy from the turbulence to the mean plasma flow, or shear flow, exceeds that being input into the turbulence itself. The plasma state is thus determined by the balance (or lack thereof) of the terms in the following equality:

\begin{equation}
    \mathcal{P} = \gamma_\mathrm{eff}E_{t}
    \label{eq:lh_general}
\end{equation}

where $\mathcal{P}$ is the production of energy from the fluctuations $\gamma_\mathrm{eff}$ is an effective growth rate for the turbulence, and $E_{t}$ is the energy in the turbulent fluctuations themselves. The term on the left-hand side (LHS), $\mathcal{P}$, acts as an energy source for the shear flow. Sometimes called the Reynolds work, it is amplified when the flow shear and the Reynolds stress are correlated, as well as when the electron pressure and potential oscillations are in phase. $\mathcal{P}$ can thus be expressed as:

\begin{equation}
    \mathcal{P} = \frac{1}{1 + \delta_{\phi,p_{e}}^{2}} \langle \tilde{u}_{x} \tilde{u}_{y} \rangle \partial_{x} \langle u_{y} \rangle
\end{equation}

where $\delta_{\phi,p_{e}}$ describes the phase between the potential and electron pressure oscillations, $\langle \tilde{u}_{x} \tilde{u}_{y} \rangle$ is the correlation between x-directed velocity fluctuations and y-directed velocity fluctuations (also known as the Reynolds stress), and $\partial_{x} \langle u_{y} \rangle$ is the x-directed derivative of the mean flow, here taken to be in the y-direction. In tokamak coordinates, the x-direction is the radial direction, and the y-direction is the binormal direction (which is normal to the magnetic field lines and has some component in both the poloidal and toroidal directions).

The term on the right-hand side (RHS) of Equation \ref{eq:lh_general} dictates the overall energy in the turbulence, which is the sum of a number of sources of free energy. Details of these mechanisms and the reasoning behind the expression for the different turbulence drives can be found in \cite{eich_separatrix_2021, manz_power_2023, grover_reduced_2024}. These can be divided into those corresponding to kinetic energy turbulence, electron free energy turbulence, and ion free energy turbulence. Thus, the RHS of Equation \ref{eq:lh_general} can be rewritten as:

\begin{equation}
    \gamma_\mathrm{eff}E_{t} = \gamma_{e}({E_{tk}} + E_{te}) + \gamma_{i}E_{ti}
\end{equation}

where $\gamma_{e}$ is the electron turbulence growth rate, related to both the energy in the kinetic energy turbulence, $E_{tk}$, and that in the electron turbulent free energy, $E_{te}$, and $\gamma_{i}$ is the ion turbulence growth rate, related to the ion turbulent free energy, $E_{ti}$.

Using substitutions for these quantities in terms of DALF-normalized parameters, Equation \ref{eq:lh_general} can be rewritten as:

\begin{equation}
    \alpha_\mathrm{RS}\frac{k_\mathrm{EM}\tau_{i}\Lambda_{pi}}{1 + (\frac{\alpha_{t}}{\alpha_{c}}k_\mathrm{EM})^{2}} = \frac{\alpha_{t}}{\alpha_{c}}(k_\mathrm{EM}^{2} + \frac{1}{2}) + \frac{1}{2k_\mathrm{EM}^{2}} \sqrt{\omega_{B}\tau_{i}\Lambda_{pi}}
    \label{eq:lh_specific}
\end{equation}

where the most influential parameters are $k_\mathrm{EM}$, the characteristic wavenumber for electromagnetic (EM) turbulence, $\alpha_{t}$, introduced earlier, $\omega_{B}$, a proxy for the magnetic curvature, and $\alpha_\mathrm{RS}$, the Reynolds stress factor. The first, $k_\mathrm{EM} = \sqrt{\frac{\beta_{e}}{\mu}}$, where $\beta_{e} = \frac{2\mu_{0}p_{e}}{B_{t}^{2}}$ and $\mu = m_{e}/m_{D}$, the electron-ion mass ratio. The expression for $\alpha_{t}$ was given in Section \ref{sec:intro}. $Z_\mathrm{eff}$, the effective plasma charge, is taken to be $Z_\mathrm{eff}$ = 1.4 for all low impurity plasmas considered here. The cuvature drive, $\omega_{B} = \frac{\lambda_{p_{e}}}{R_\mathrm{geo}}$, where $\lambda_{p_{e}}$ is chosen as the perpendicular wave number, $\lambda_{\perp}$, and is related to the growth rate of the ideal interchange instability, taken here to be the growth rate of the ion turbulence term. Discussion of $\alpha_\mathrm{RS}$ will be deferred to Section \ref{sec:reverse_field_os}, but for all plasmas in the favorable drift direction considered in this section, $\alpha_\mathrm{RS} = 1$.

Also in Equation \ref{eq:lh_specific} are $\alpha_{c}$, $\tau_{i}$, and $\Lambda_{pi}$. $\alpha_{c}$ denotes the critical normalized pressure gradient and is related to the parallel wave number, $k_{\parallel}$, given in terms of shaping parameters. It does not vary substantially between L- and H-modes. Its definition can be found in \cite{eich_correlation_2018}. Finally, $\tau_{i} = \frac{T_{i}}{T_{e}}$ is the ratio of the ion to electron temperatures, and $\Lambda_{pi} = \frac{\lambda_{p_{i}}}{\lambda_{p_{e}}}$ is the ratio of the ion to electron pressure gradient scale lengths. Lack of availability of ion measurements near the separatrix makes measuring these quantities challenging, compelling the choice of simply setting their product, $\tau_{i}\Lambda_{pi}$ = 1. The left panel of Figure \ref{fig:sepos_norm} shows the L-H criterion in normalized units for the superset, corresponding to the full range of parameters in Table \ref{tab:datasets}. The stabilizing term (LHS of Equation \ref{eq:lh_specific}) is shown on the x-axis and the destabilizing term (RHS of Equation \ref{eq:lh_specific}) is shown on the y-axis. Even for this larger dataset without sub-selected $\mathcal{M}$, the balance $y = x$ line does a good job of separating the L-modes from the H-modes. 


\subsection{L-mode density limit}

The LDL has been observed experimentally on many devices and presents a fundamental limit to tokamak operation in the L-mode \cite{greenwald_new_1988, petrie_plasma_1993, greenwald_density_2002}. It is thought that an increase in density beyond allowable levels increases edge collisionality, driving resistive ballooning mode (RBM) transport \cite{rogers_phase_1998}. Electromagnetic fluctuations then lead to further destabilization of RBM fluctuations, leading to the collapse of the plasma column observed in LDL-driven disruptions \cite{rogers_enhancement_1997, eich_turbulence_2020, manz_power_2023}. Inspired by these observations, the SepOS model proposes that both the previously introduced $k_\mathrm{EM}$, as well as the wavenumber characteristic of RBM turbulence, $k_\mathrm{RBM}$, play important roles in this collapse. In particular, the LDL is triggered when the following is satisfied:

 \begin{equation}
     k_\mathrm{EM} = k_\mathrm{RBM}
     \label{eq:LDL}
\end{equation}

where $k_\mathrm{RBM} = \frac{k_{\parallel}}{\sqrt{(1 + \tau_{i}) C \omega_{B}}}$, with $k_{\parallel} = \sqrt{\alpha_{c}}$, the critical value of the normalized pressure gradient, $\alpha_\mathrm{MHD}$, and $C$ is a parameter introduced in \cite{scott_nonlinear_2002}, related to $\alpha_{t}$ through the pressure gradient in $\omega_{B}$. Importantly, Equation \ref{eq:LDL} is a condition for the \emph{L-mode} density limit. In H-mode, the condition can be met, and as long as the LHS of equations \ref{eq:lh_general} or \ref{eq:lh_specific} is larger than the RHS, the plasma will remain in a stable H-mode. If an H-L back-transition occurs (RHS $>$ LHS of Equations \ref{eq:lh_general} or \ref{eq:lh_specific}), a plasma will either disrupt if $n_{e} > n_{e}^\mathrm{LDL}$ or remain a stable L-mode if $n_{e} < n_{e}^\mathrm{LDL}$ for a particular $T_{e}$.

Due to difficulties with robustly measuring $\lambda_{T_{e}}$ and other associated separatrix quantities at the high $n_{e}$ required to trigger DLs on C-Mod, data for DLs are not available for comparison for this particular exercise. Figure \ref{fig:ff_sepos} builds confidence that for L-mode discharges across a moderate range of parameter space (even restricted by the selection of $\mathcal{M}$), all but two L-modes with the largest error bars fall to the left of the LDL boundary. The center panel of Figure \ref{fig:sepos_norm} shows the larger dataset across a range of $\mathcal{M}$ in normalized variables and as with the sub-selected dataset, most L-mode discharges fall below the LDL, i.e. $k_\mathrm{EM} > k_\mathrm{RBM}$. The figure also shows that many H-modes do have $k_\mathrm{EM} < k_\mathrm{RBM}$, yet do not disrupt.


\subsection{Ideal ballooning MHD limit}

H-modes, however, are subject to the third limit described in the SepOS model, the IBML, which occurs in the transition between resistive to ideal MHD turbulence. The separatrix becomes unstable to ideal MHD ballooning when the following condition is met:

 \begin{equation}
     k_\mathrm{ideal} = k_\mathrm{RBM}
     \label{eq:IBML}
\end{equation}


where $k_\mathrm{ideal} = \sqrt{\frac{\beta_{e}\sqrt{\omega_{B}}}{C}}\frac{q_{s}R}{\lambda_{\perp}}$, with $q_{s}$, the safety factor, taken to be $\hat{q}_\mathrm{cyl}$. As with the LDL, this limit is presented as a balance of wavenumbers, rather than of energy transfer as was the case for the L-H criterion. The right panel of Figure \ref{fig:sepos_norm} shows that all L-mode and most H-mode discharges fall under this balance line. Unlike the LDL, which only applies to L-modes, the IBML applies to both, namely because a discharge with Equation \ref{eq:lh_specific} categorizing it as an L-mode will also have $k_\mathrm{RBM} < k_\mathrm{ideal}$. Equation \ref{eq:IBML} is essentially a recasting of the single fluid ideal MHD limit to the normalized pressure gradient, $\alpha_\mathrm{MHD}$, given by: 

 \begin{equation}
     \alpha_\mathrm{MHD} = \alpha_{c}
     \label{eq:IBML_alpha}
\end{equation}

where $\alpha_\mathrm{MHD} = R_\mathrm{geo} q_\mathrm{cyl}^{2} \frac{\beta_{e}}{\lambda_{p}}$. More details on the connection between Equations \ref{eq:IBML} and \ref{eq:IBML_alpha} and how $\alpha_\mathrm{MHD}$ varies as discharges approach the boundary on AUG can be found in \cite{eich_correlation_2018, eich_separatrix_2021}.

\section{SepOS in the unfavorable drift direction}
\label{sec:reverse_field_os}

The next dataset to be analyzed contains is given by the third column in Table \ref{tab:datasets} and contains only discharges in the unfavorable $\nabla B$-drift direction, i.e. when the ion $\nabla B$-drift points opposite to the active X-point. As noted above, while the LDL and IBML are not thought to depend on drift direction, the L-H transition does  \cite{ryter_h-mode_1995, hubbard_local_1998, carlstrom_comparison_1998}. Running in the unfavorable drift direction also offers the advantage of facilitating access to the so-called ``improved" confinement mode (I-mode) \cite{whyte_i-mode_2010, hubbard_edge_2011}. Indeed many discharges in this second dataset have been identified as I-modes and application of the SepOS to these allows the beginning of a survey of I-mode access from the separatrix perspective. Like the EDA H-mode, I-mode is a regime free of large ELMs and offers improved confinement relative to the L-mode. As such, understanding I-mode access is key to evaluating its potential as a reactor-relevant scenario.

The SepOS framework has been recently applied to H-modes in the unfavorable drift direction on AUG \cite{grover_reduced_2024}. In line with the expected increase in $P_\mathrm{SOL}$ required to reach the L-H transition, that analysis found that for the same value of $n_{e}^\mathrm{sep}$, H-modes appeared at higher $T_\mathrm{sep}$. Across the dataset, this corresponded loosely to an upwards shift of the L-H transition curve. This shift was understood in the context of the SepOS model through the introduction of the proportionality constant, $\alpha_\mathrm{RS}$, in the transfer of energy between the Reynolds stress and the average radial fluctuations energy, consistent with observations in \cite{cziegler_turbulence_2017}. This constant modifies the stabilizing term in the L-H transition criterion, entering the LHS of Equation \ref{eq:lh_specific}. It describes the average tilt of the turbulent eddies, which depends on the flow shear, as well as the magnetic shear, and is based on the eddy-tilting picture studied in \cite{fedorczak_shear-induced_2012, fedorczak_dynamics_2013}. In the favorable drift direction, these components point in the same direction, and constructively add to give $\alpha_\mathrm{RS}$ = 1. In the unfavorable drift direction, however, they point in opposite directions, and add to yield a value of $\alpha_\mathrm{RS} <$ 1. From a set of gyrokinetic simulations, the study on AUG found that a value of $\alpha_\mathrm{RS}$ = 0.4 well-described the separation between unfavorable drift L-modes and H-modes across a sizable variation in $B_{t}$, $I_{P}$, and $\hat{q}_\mathrm{cyl}$.



\begin{figure}[h]
\centering
\includegraphics[width=\columnwidth]{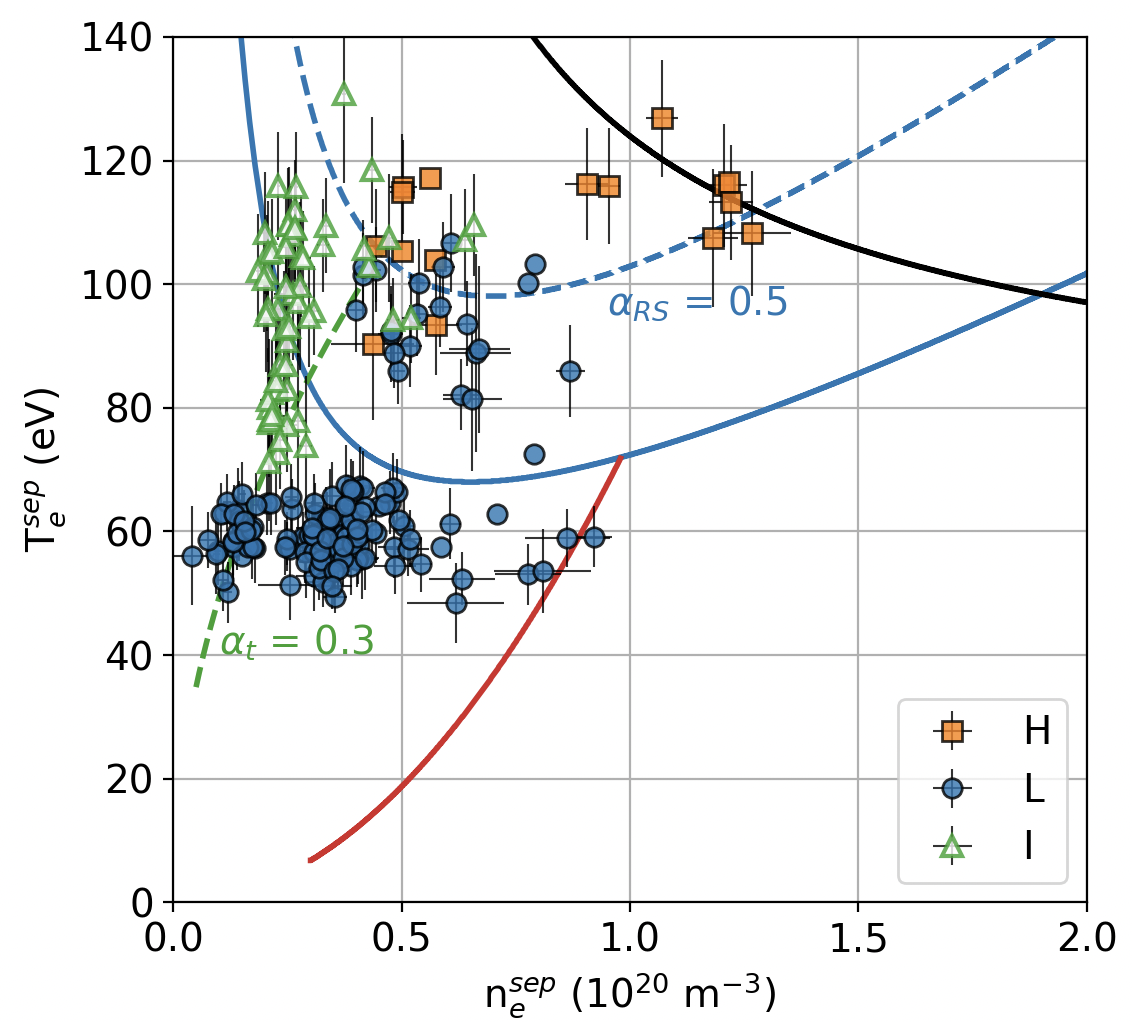}
\caption{Separatrix operational space in the unfavorable drift direction, in terms of $n_{e}$ and $T_{e}$ for the typical magnetic equilibrium parameters from Table \ref{tab:mag_geo}. H-modes are orange squares, L-modes are blue circles, and now I-modes are open green triangles. The three main SepOS curves are shown as well, closely matching those in Figure \ref{fig:ff_sepos}. Shown also is a the contour of $\alpha_{t}$ = 0.3 and the L-H curve corrected with $\alpha_\mathrm{RS}$ = 0.5 as dotted lines in green and blue, respectively. For reference the curve at $\alpha_\mathrm{RS}$ = 1 is shown as well.}
\label{fig:rf_sepos}
\end{figure}

\begin{figure*}[h]
\centering
\includegraphics[width=1.6\columnwidth]{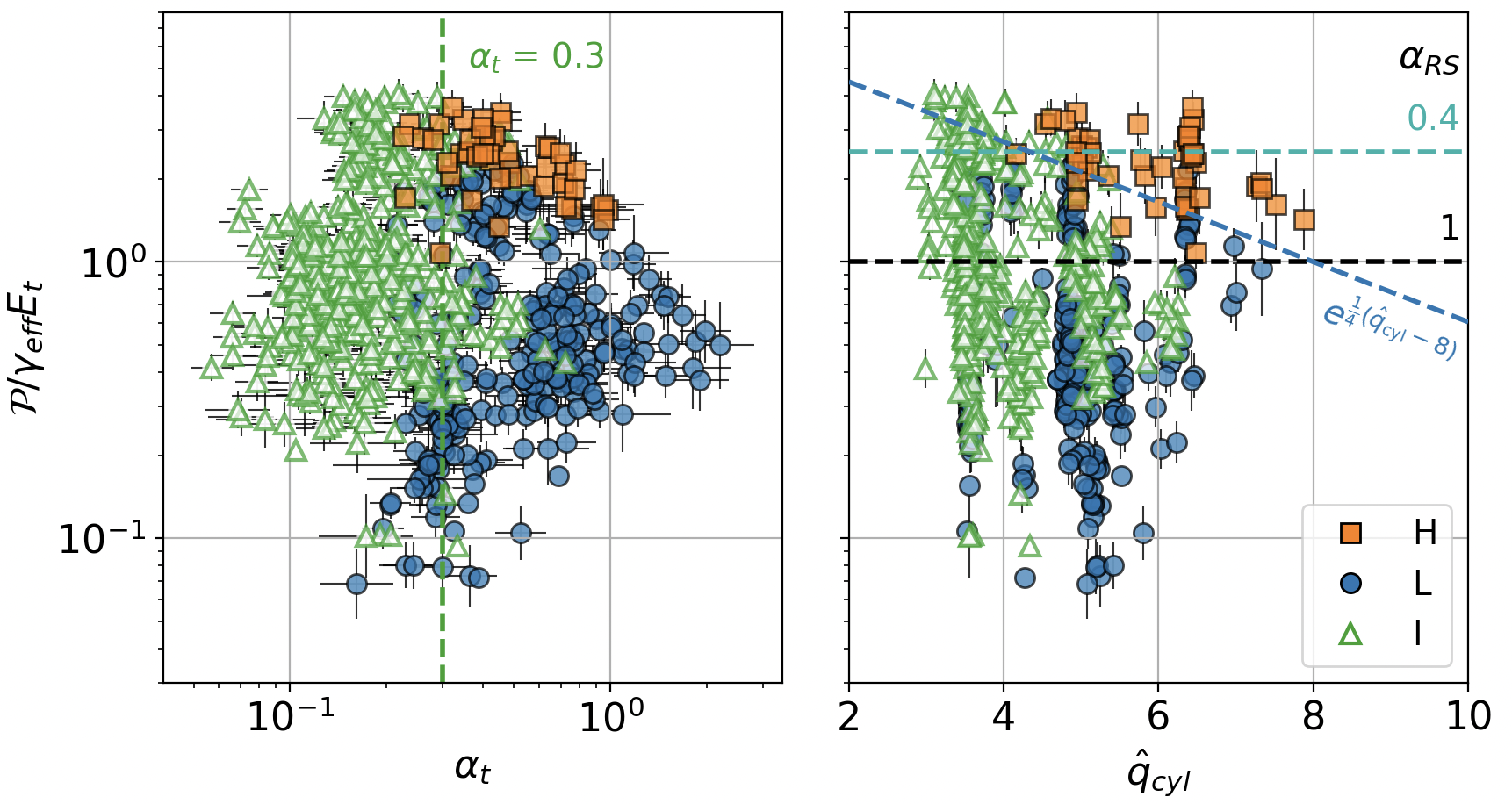}
\caption{Ratio of Reynolds energy transfer rate to turbulent energy input rate plotted against $\alpha_{t}$ (left) and $\hat{q}_\mathrm{cyl}$ (right). Color coding of symbols is the same as that in Figure \ref{fig:rf_sepos}. The dashed vertical green line in the plot at left shows the fixed $\alpha_{t}$ = 0.3 line. At right, dashed lines represent different functional forms for $\alpha_\mathrm{RS}$ in the unfavorable drift direction. The horizontal black line and horizontal turquoise line show values of $\alpha_\mathrm{RS}$ = 1 and 0.4 respectively. The dark blue line shows a manual fit to $\alpha_\mathrm{RS} = \alpha_\mathrm{RS}(\hat{q}_\mathrm{cyl})$, designed to separate L/I- modes from H-modes using an exponential fit (linear in log space).}
\label{fig:rf_lih}
\end{figure*}

Figure \ref{fig:rf_sepos} shows the separatrix operational space on C-Mod for discharges in the unfavorable drift direction with the same $B_{t}$ = 5.4 T and $I_{P}$ = 0.8 MA as that in figure \ref{fig:ff_sepos}. The full range in parameters for this typical operating scenario is the same as that listed in Table \ref{tab:mag_geo}. The figure shows, when compared against Figure \ref{fig:ff_sepos}, that L-modes exist at temperatures higher than in the favorable drift direction, even above 100 eV. While there is a less clear separation between L- and H-modes than in discharges in the favorable drift direction, it is clear that some modification to the L-H criterion is required. Including the effect of weaker stabilization via the introduction of a value for $\alpha_\mathrm{RS} <$ 1, however, allows for better description of the transition with the different $\nabla B$-drift direction. For this selection of $\mathcal{M}$, a value of $\alpha_\mathrm{RS} = 0.5$, close to that on AUG, best separates the L- and H-modes, although some L-modes and some H-modes fall on the incorrect side of the boundary.

Figure \ref{fig:rf_sepos} also includes I-mode discharges. These span a large range of $T_{e}^\mathrm{sep}$, but tend to have low $n_{e}^\mathrm{sep}$ and thus generally lie to the left of both L-mode and H-mode discharges, for fixed $T_{e}^\mathrm{sep}$. This is consistent with the finding on AUG that I-modes exist at low separatrix density, with $n_\mathrm{sep}/n_\mathrm{GW} < 0.25$ \cite{silvagni_pellet-fueled_2023}, where $n_\mathrm{GW}$ is the Greenwald fraction. Notably, most I-modes fall on the L-mode side of the L-H curve, when including the modification of the criterion via $\alpha_\mathrm{RS}$ = 0.5. I-modes are observed to exhibit suppressed heat transport, but not particle transport \cite{hubbard_edge_2011}, and so it is intuitive that they might not experience as large of a transfer of turbulent energy into shear flow as an H-mode. Figure \ref{fig:rf_sepos} also shows a contour of constant $\alpha_{t}$ = 0.3. For this selection of $\mathcal{M}$, most I-modes fall to the left of this contour. I-modes are typically observed at low edge collisionalities \cite{hubbard_multi-device_2016, hubbard_physics_2017, manz_physical_2020}, so a ceiling on $\alpha_{t}$ limiting I-mode access is not surprising. Regardless, work to develop a more robust criterion for I-mode sustainment based on DALF turbulence is essential for projection and is ongoing.

Figure \ref{fig:rf_lih} shows the entire dataset in the unfavorable drift direction, using the same denomination for L-modes, H-modes, and I-modes as in Figure \ref{fig:rf_sepos}. The figure shows the ratio of the zonal flow production term, $\mathcal{P}$, to the turbulent energy input, $\gamma_\mathrm{eff}E_{t}$, from Equation \ref{eq:lh_general}, as a function of $\alpha_{t}$ and $\hat{q}_\mathrm{cyl}$. Note that the these plots do not include any correction to the $\mathcal{P}$ term in Equation \ref{eq:lh_general} via $\alpha_\mathrm{RS}$. As in Figure \ref{fig:rf_sepos}, however, it is clear that some $\alpha_\mathrm{RS} < 1$ correction is required to properly classify L- and H-modes. The plot on the left of Figure \ref{fig:rf_lih} confirms that across the dataset on C-Mod, I-modes exist at lower values of $\alpha_{t}$ than both H-modes and L-modes. Although not a perfect separator, a value of $\alpha_{t}$ = 0.3 appears to adequately distinguish I-modes from both L- and H-modes. 

As mentioned earlier, it may be expected that I-modes, as a result of insufficient energy transfer to zonal flows, should fall on the same side of the L-H transition as L-modes. In some sense, this would confirm that the criterion in Equation \ref{eq:lh_general} is one of H-mode access/sustainment, rather than of the L-H transition strictly. When considering the plot on the right of Figure \ref{fig:rf_lih}, I-modes do tend to be below H-modes for moderate to high values of $\hat{q}_\mathrm{cyl}$, i.e. the ratio of $\mathcal{P}/\gamma_\mathrm{eff}E_{t}$ for I-modes is more similar to that of L-modes than H-modes. Plotting against $\hat{q}_\mathrm{cyl}$, however, reveals that the boundary between H-modes and L/I-modes may vary with $\hat{q}_\mathrm{cyl}$. At low values of $\hat{q}_\mathrm{cyl}$, a ratio of $\mathcal{P}/\gamma_\mathrm{eff}E_{t} = 0.3 - 0.4$ may be appropriate to keep all I-modes below the boundary, whereas at high $\hat{q}_\mathrm{cyl}$, the ratio approaches 1. This might suggest that the proportionality constant, $\alpha_\mathrm{RS}$, may have some dependence on $\hat{q}_\mathrm{cyl}$. Given that $\alpha_\mathrm{RS}$ is a metric for the relative contributions of magnetic shear and flow shear, it may be the case that this value is dependent on the strength of the magnetic shear itself, given by $\hat{q}_\mathrm{cyl}$. The right plot of Figure \ref{fig:rf_lih} shows increased appearance of I-modes (and inhibition of H-modes) at low $\hat{q}_\mathrm{cyl}$, and harder access to I-modes (easier access to H-modes) at high $\hat{q}_\mathrm{cyl}$. This is consistent with control room reports of easier access to I-mode at high $I_{P}$, and may be connected with the sudden vs. gradual L-I transitions reported at low and high $q_{95}$ in \cite{whyte_i-mode_2010}. A manually chosen boundary, imposing dependence on $\hat{q}_\mathrm{cyl}$ to the L/I-H transitions, is also shown. Note that the curve is intended only to separate confinement modes for $\mathcal{P}/\gamma_\mathrm{eff}E_{t} > 1$, and is not meant to imply  $\alpha_\mathrm{RS} > 1$ and easier H-mode access in the unfavorable drift-direction at high $\hat{q}_\mathrm{cyl}$ than in the favorable drift-direction. 

The I-mode preference for low edge collisionality and safety factor is apparent from Figure \ref{fig:rf_lih}. What is not immediately obvious from this dataset alone is whether the lack of H-modes at low $\hat{q}_\mathrm{cyl}$ (and possibly L-modes at high $\hat{q}_\mathrm{cyl}$) is a limitation of the current dataset or a comment on the physics behind the L-I-H transitions. Work to expand this dataset to include more unfavorable L-modes and H-modes is underway. In the absence of an expanded dataset, the crude fit in Figure \ref{fig:rf_lih} and identification of a possible threshold in $\alpha_{t}$ at least qualitatively points to dependence of $\hat{q}_\mathrm{cyl}$ in the the L-I-H transitions in the unfavorable drift direction.

\section{Transition between ELMy and EDA H-mode discharges}
\label{sec:elmy_eda}

The intrinsic ELM-free nature of the I-mode makes it an attractive candidate for further study, as it is expected that few, if any, large ELMs (usually called ``Type-I'') will be tolerable in next-generation devices \cite{leonard_impact_1999, hughes_power_2011, kuang_divertor_2020}. As for H-modes, it has been observed that the type of H-mode, including those without Type-I ELMs, is associated with differing values of edge $n_{e}$ and $T_{e}$ independently, and more specifically with $\nu^{*}$ and $\alpha_\mathrm{MHD}$ \cite{hubbard_h-mode_2007, hughes_edge_2007, hughes_pedestal_2013}. While it is not clear where in the edge these differences are most important, this section considers the separatrix to assess access conditions for different H-mode types.

\begin{figure*}[h]
\centering
\includegraphics[width=1.5\columnwidth]{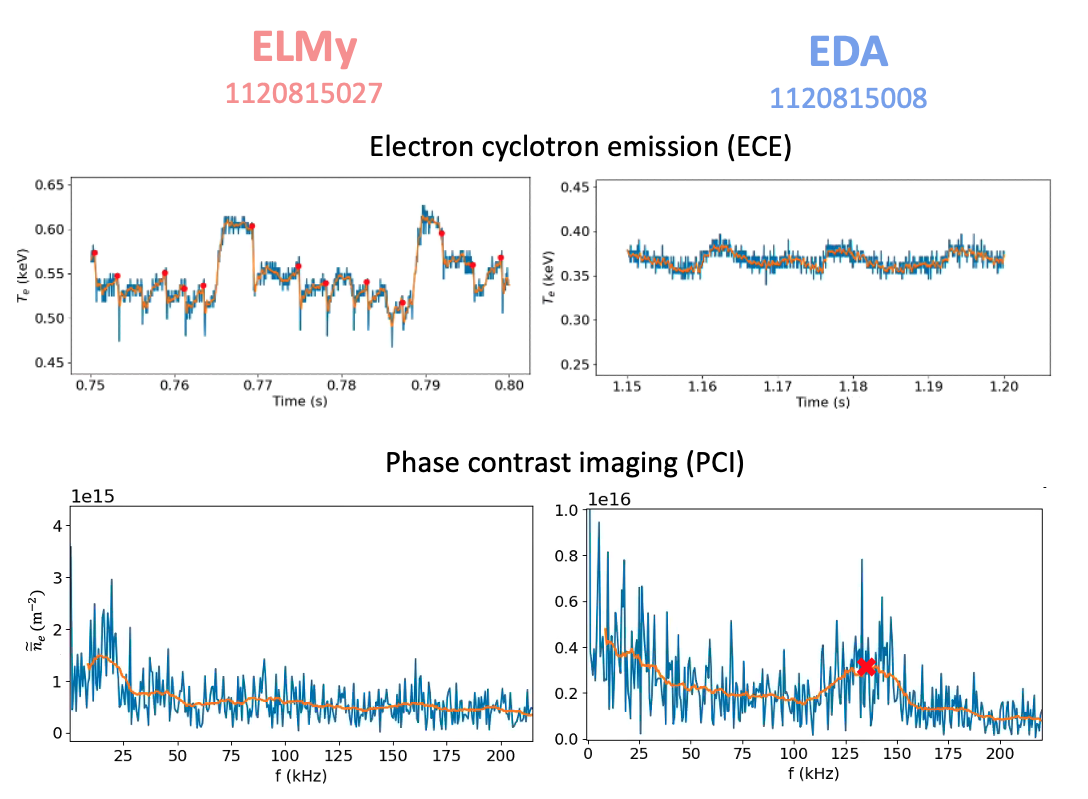}
\caption{Measurements from outermost channel of the electron cyclotron emission (top) and the 15th chord of the phase contrast imaging (bottom). Measurements are shown for time slices identified as ELMy (left) and EDA (right). Blue lines show the raw signals and orange lines show a moving average to the signal to allow for easier identification of ELMs and the QCM. ELMs are identified by sharp drops in $T_{e}$ ($\frac{\partial T_{e}}{\partial t} < -100$ keV/m) near the pedestal top and the start of ELM crashes is denoted by red circles. The QCM is identified in fluctuations of line-averaged density by searching for a coherent peak in the power spectrum somewhere between 50 -- 150 kHz and the mode height is denoted by a red cross.}
\label{fig:elmy_eda_id}
\end{figure*}

In particular, this work focuses on the enhanced D$_{\alpha}$ (EDA) regime \cite{greenwald_characterization_1999}, closely related to the quasi-continuous exhaust (QCE) regime \cite{faitsch_broadening_2021}, both of which are free of Type-I ELMs and are favored by similar plasma conditions. These H-modes are found to have a higher pedestal density, $n_{e}^\mathrm{ped}$, and lower pedestal temperature, $T_{e}^\mathrm{ped}$, i.e. a higher pedestal collisionality, $\nu^{*}_\mathrm{ped}$. While differences between the EDA and the QCE regime are still under investigation, they are both high collisionality modes that are characterized by the appearance of a quasi-coherent mode (QCM) near the bottom of the pedestal \cite{snipes_quasi-coherent_2001, golfinopoulos_edge_2018, kalis_experimental_2024}. The mode is thought to drive increased transport in the edge \cite{terry_transport_2005}, preventing the pedestal pressure gradient and height from reaching the peeling-ballooning limit associated with Type-I ELMs \cite{hughes_pedestal_2013}. This increased transport is associated with enhanced D$_{\alpha}$ emission in the edge in the case of the EDA and enhanced filamentary transport in the form of small ELMs in the case of the QCE. rk on AUG has indicated that  at the separatrix there is a clear separation between the Type-I ELMy and QCE regimes and that the separation is best described by $\alpha_{t}$ \cite{faitsch_analysis_2023}.

To this end, the transition between the Type-I ELMy and the EDA H-mode on C-Mod is scrutinized using the SepOS framework. Type-I ELMs were not frequently observed on Alcator C-Mod \cite{hughes_pedestal_2013}. Operation at high $n_{e}$ yielded high edge $\nu^{*}$ \cite{hughes_observations_2002}, more compatible with the EDA regime. Type-I ELMs were not produced until later in its operation and were found to be most easily produced in a significantly different magnetic configuration to that of the typical operating scenario \cite{TERRY2007994}. A more closed divertor and weaker shaping helped facilitate Type-I ELMy H-mode access. One particular experiment performed in 2012 sought to study the transition between the EDA H-mode and the ELMy H-mode in the same shape. It made use of a suite of fluctuation measurements to characterize changes to the turbulence in the two regimes \cite{diallo_observation_2014, diallo_quasi-coherent_2015}. 

Experimentally, the transition from EDA to ELMy H-mode was stimulated by reducing the L-mode target density \cite{diallo_quasi-coherent_2015}. The run day began with high density EDA H-modes and as the day progressed, the target density was gradually reduced until at a critical value, type-I ELMs emerged. Transitions between the two regimes, however, sometimes occurred during shots, and it is not possible to label a shot as purely EDA or purely ELMy. Instead, stationary periods are broken up into 50 ms phases, each of which receive either the ``EDA" or ``ELMy" label, according to a set of criteria. 

Figure \ref{fig:elmy_eda_id} shows an example of this labeling procedure. EDA discharges are identified using measurements from the phase contrast imaging (PCI) diagnostic. The PCI on C-Mod measured fluctuations in the line-averaged density, $\tilde{\overline{n}}_{e}$, along 32 vertical chords \cite{mazurenko_experimental_2002, golfinopoulos_external_2014}. The PCI power spectrum is computed using a fast Fourier transform, and a coherent mode ranging between 50 -- 150 kHz is sought in this spectrum for each chord. If the mode is found in this frequency range for 80\% of the chords, the time window is identified as an ``EDA". For identification of ELMy discharges, the electron cyclotron emission (ECE) diagnostic is used -- in particular, the last channel of a nine channel grating polychromator (GPC) \cite{hubbard_measurements_1998}. The GPC made highly time-resolved measurements of $T_{e}$ on C-Mod, and its outermost channel well characterized $T_{e}$ near the pedestal top, which enabled ELM cycle resolution. A discharge is categorized as ``ELMy" if the time derivative of $T_{e}$ from GPC, $\frac{\partial T_{e}}{\partial t} < -100$ keV/m. This workflow and criterion is similar to that developed in \cite{hughes_pedestal_2013}. For the purposes of this work, no distinction is made between large and small ELMs, but the criterion in $\frac{\partial T_{e}}{\partial t}$ ensures a large enough change in the pedestal to exclude very small ELMs.

\begin{figure}[h]
\centering
\includegraphics[width=\columnwidth]{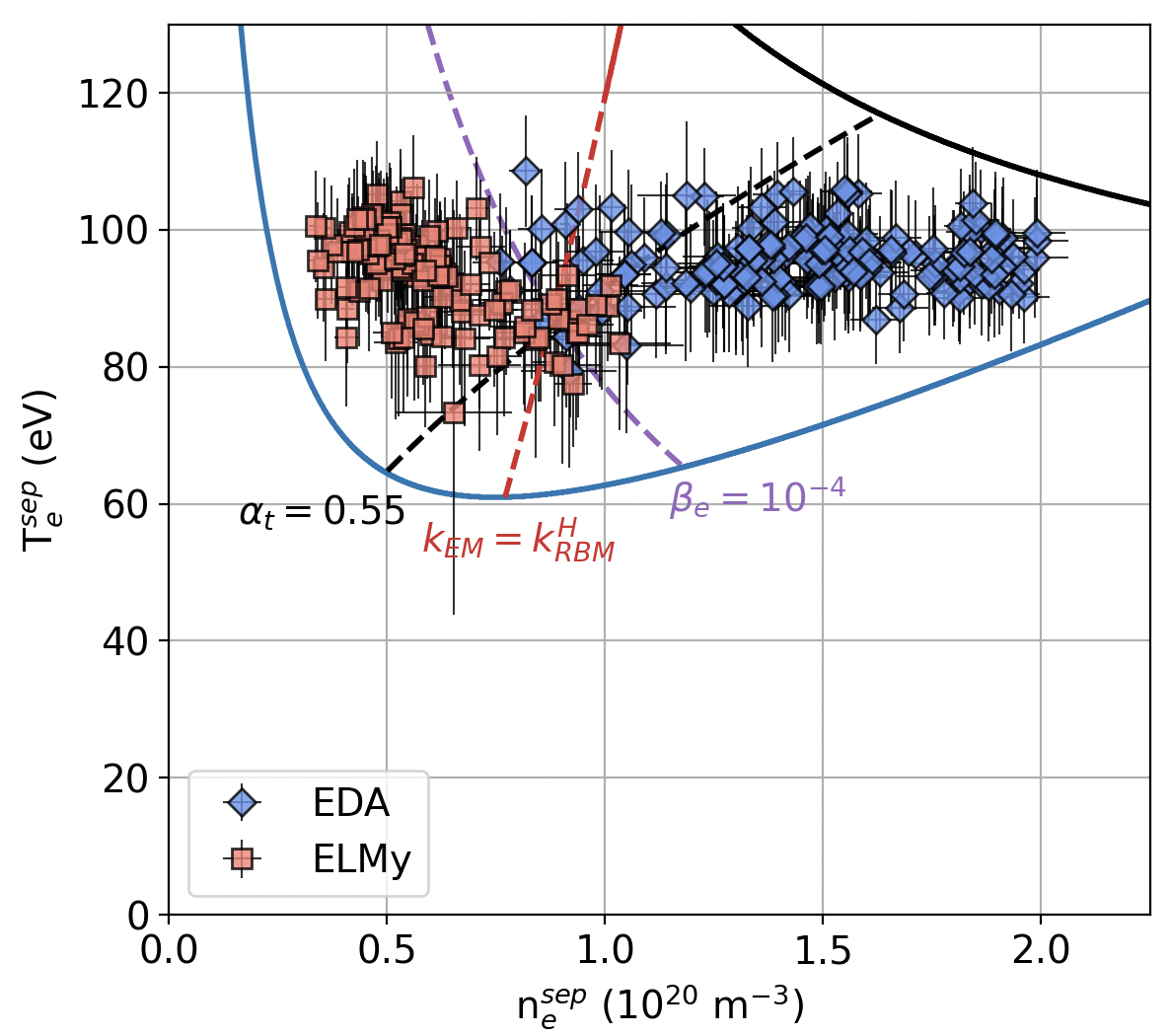}
\caption{The separatrix operational space for the experiment detailed in \cite{diallo_observation_2014, diallo_quasi-coherent_2015} in terms of $n_{e}$ and $T_{e}$. Points are categorized as EDA (blue diamonds) or ELMy (red squares). The L-H and IBML boundaries are shown in blue and black, respectively. The three dashed curves bisecting the H-mode space show $\alpha_{t}$ = 0.55 in black, $\beta_{e}$ = $10^{-4}$, in purple, and $k_\mathrm{EM} = k_\mathrm{RBM}^{H}$ in red.}
\label{fig:elmy_eda_sepos}
\end{figure}

Having performed this categorization, $n_{e}^\mathrm{sep}$ and $T_{e}^\mathrm{sep}$ are identified for each 50 ms phase, and the results are plotted in Figure \ref{fig:elmy_eda_sepos}. As an additional check on Equations \ref{eq:lh_general} and \ref{eq:lh_specific} of the SepOS model, it can be seen that all data points lie above the L-H boundary. Note also that this dataset includes no L-mode phases, focusing exclusively on the H-mode phase. The same methodology as in Section \ref{sec:lh} is used, albeit for a different $\mathcal{M}$, resulting from the different shape (although $I_{P}$ and $B_{t}$ are similar to that shown in Figure \ref{fig:ff_sepos}). The figure shows variation of $T_{e}^\mathrm{sep}$ of less than 40 eV, while $n_{e}^\mathrm{sep}$ varies from just above 0.3 $\times 10^{20}$ m$^{-3}$ to just under 2.0 $\times 10^{20}$ m$^{-3}$. The EDA H-mode in particular appears to see very little variation in $T_{e}^\mathrm{sep}$ for large changes in $n_{e}^\mathrm{sep}$. Whether the large variation in $n_{e}^\mathrm{sep}$ at relatively invariant $T_{e}^\mathrm{sep}$ is a feature of the EDA regime or rather a byproduct of the experimental setup is not immediately clear. A recent study of an effective scan in $P_\mathrm{SOL}$ for a dataset of EDA H-modes on C-Mod, however, similarly observed large changes in $n_{e}^\mathrm{sep}$ for much smaller changes in $T_{e}^\mathrm{sep}$ \cite{miller_collisionality_2024}. That study found that despite experimental changes to the heat sources at the separatrix, the changes in the plasma were linked an increase in the particle source. Ionization was found to be tightly coupled to variation in $P_\mathrm{SOL}$, resulting in large changes to $n_{e}^\mathrm{sep}$ specifically. A similar modification of the ionization source is likely at play in this work, but as a result of direct changes to fueling, rather than $P_\mathrm{SOL}$, given the experimental setup.

In addition to showing the L-H transition curve, Figure \ref{fig:elmy_eda_sepos} shows three other boundaries, each of which represent potential transitions between the two H-mode regimes. The first of these follows from work on AUG demonstrating that Type-I ELMy H-modes and QCE discharges were well separated by a constant value of $\alpha_{t} = 0.55$ \cite{faitsch_analysis_2023}. The dashed black curve shows this value of $\alpha_{t}$ and indeed, most ELMy discharges lie at $\alpha_{t} < 0.55$ and most EDA discharges (similar to QCE) lie at $\alpha_{t} > 0.55$. A second parameter that does a similarly good job of separating out the regimes is $\beta_{e}$, or equivalently $p_{e}$ (different by a factor of $B_{t}^{2}/2\mu_{0}$, which is nearly constant in this dataset). In particular, a value of $\beta_{e} = 10^{-4}$, plotted as a dashed purple line, provides good separation between the two regimes. From the figure, the constraints on $\alpha_{t}$ and $\beta_{e}$ appear to serve as bounds for which a discharge will be solidly an ELMy H-mode or solidly an EDA discharge. The region within 10 eV and 0.2 $\times 10^{20} m^{-3}$ about the intersection of $\alpha_{t} = 0.55$ and $\beta_{e} = 10^{-4}$ appears to characterize an overlap region, where either an EDA or an ELMy H-mode are possible.

\begin{figure}[h]
\centering
\includegraphics[width=0.9\columnwidth]{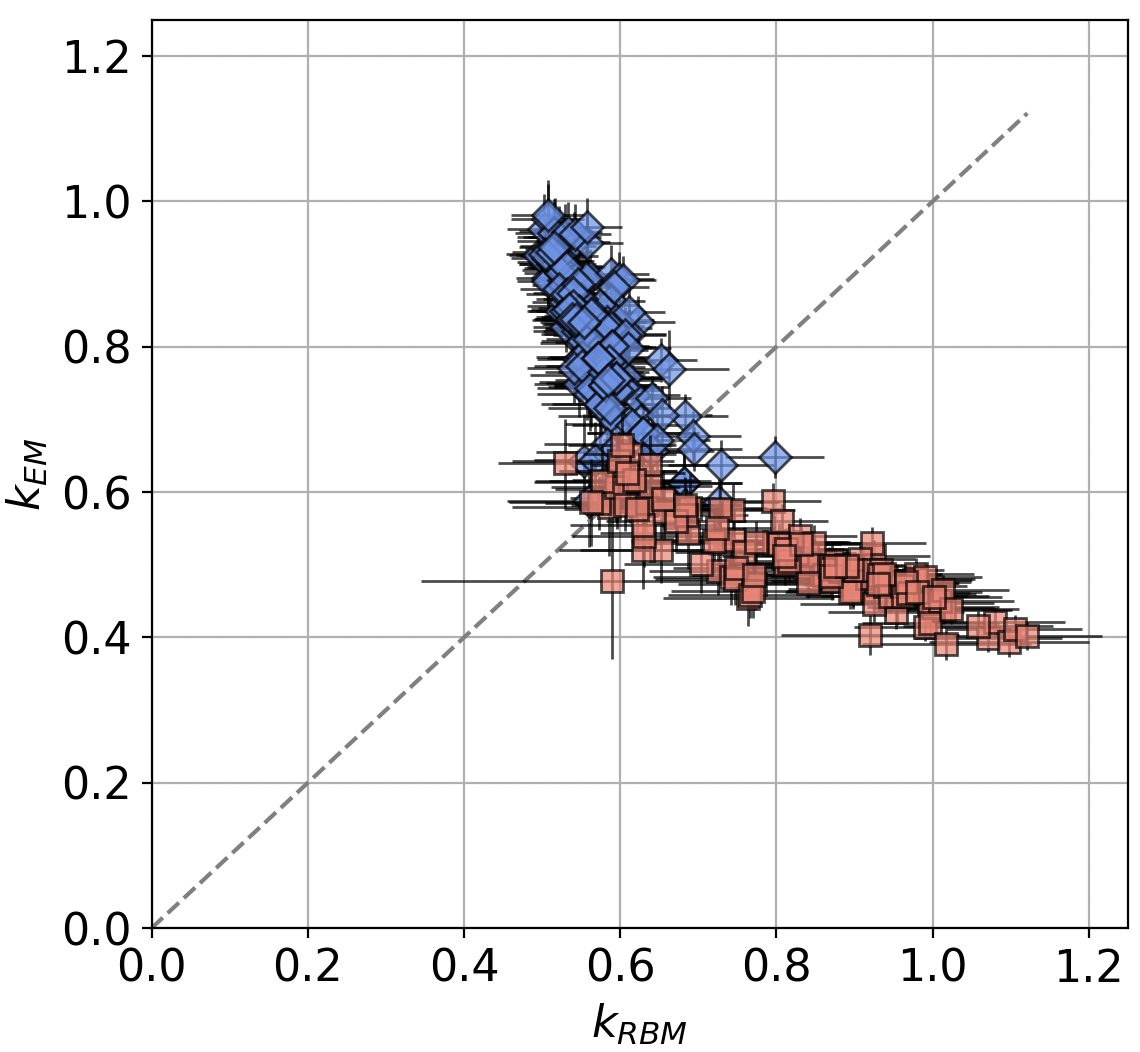}
\caption{Wavenumber for electromagnetic fluctuations plotted against that for fluctuations from the resistive ballooning mode. Points use the same color scheme as in figure \ref{fig:elmy_eda_sepos}. The equality line is shown as a dashed gray line.}
\label{fig:kEM_kRBM}
\end{figure}

One final curve is plotted in Figure \ref{fig:elmy_eda_sepos}, plotted in red -- the curve along which $k_\mathrm{EM} = k_\mathrm{RBM}$. This curve is the extension of the red curve in Figure \ref{fig:ff_sepos}, characterizing the LDL. Now it goes through the H-mode region, and uses $k_\mathrm{RBM}^{H}$, constructed from the the H-mode scaling for $\lambda_{p}$ introduced in Section \ref{sec:thomson}. As mentioned earlier, H-modes with $k_\mathrm{RBM} < k_\mathrm{EM}$ do not undergo a DL disruption, but instead, largely appear as stable EDA H-modes. While the red curve does not as cleanly separate either all ELMy or all EDA discharges, it does intersect the discharges almost perfectly in the middle of both sets. Figure \ref{fig:kEM_kRBM} shows this split more clearly in dimensionless space. Interestingly, the unity line in this plot, in addition to separating discharges identified as ELMy and EDA, also passes through a fairly clear break in slope in the trend of $k_\mathrm{EM}$ with $k_\mathrm{RBM}$. This equality of wavenumbers may indicate that when RBM turbulence evolves to scales comparable to that of EM turbulence, \emph{and} there is enough Reynolds work transferring turbulent energy into zonal flow, as per Equation \ref{eq:lh_general}, the result is enhanced transport, the appearance of coherent fluctuations in the form of the QCM and/or filamentary transport. And importantly, it results in the disappearance of large ELMs.

\begin{figure}[h]
\centering
\includegraphics[width=\columnwidth]{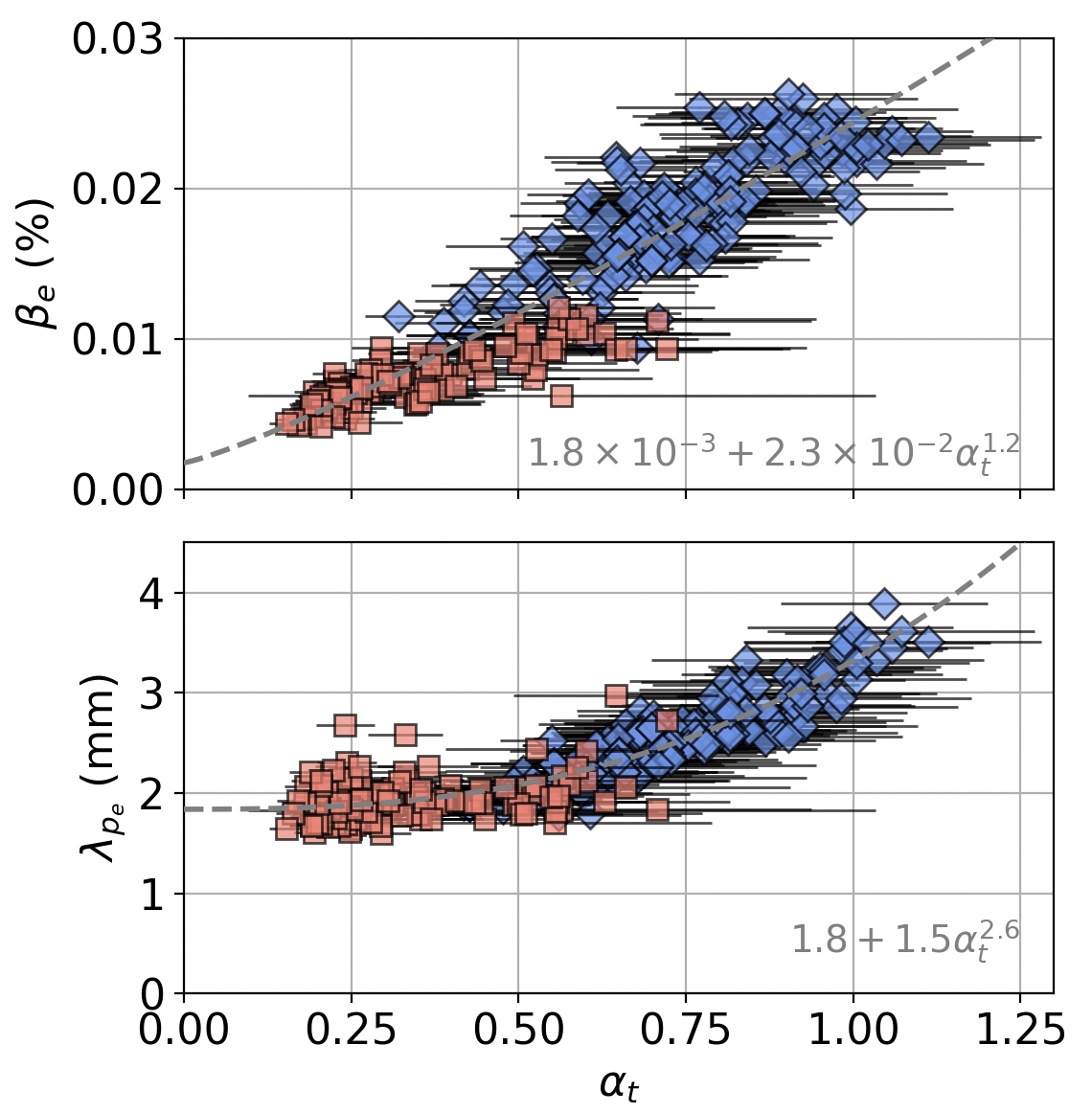}
\caption{Plasma beta (top) and electron pressure gradient scale length (bottom) plotted against $\alpha_{t}$. Points use the same color scheme as in Figure \ref{fig:elmy_eda_sepos}. Shown also are dashed gray lines, which represent nonlinear least squares fits to the data. The fit form is similar to that from Equation \ref{eq:lambdap_widening}, but excludes dependence on $\rho_{s,p}$, since this dataset features little variation in $B_{p}$.}
\label{fig:alpha_beta_lambda}
\end{figure}

That these three curves seem to all adequately describe this transition may not entirely be a coincidence. Equation \ref{eq:LDL} can be rewritten in terms of its constituent parameters as follows:

\begin{equation}
    \frac{\alpha_{t}\beta_{e}}{\sqrt{\lambda_{p_{e}}}} = \alpha_{c}\mu \sqrt{\frac{2}{R_\mathrm{geo}}}
    \label{eq:elmy_eda}
\end{equation}

where the term on the RHS of Equation \ref{eq:elmy_eda} is relatively fixed for a given magnetic configuration, $\mathcal{M}$, with $\alpha_{c}\mu\sqrt{\frac{2}{R_\mathrm{geo}}} = (1.4 \pm 0.02) \times 10^{-6}$ m$^{-1/2}$ for this dataset. The terms on the LHS of the equation all vary considerably for the discharges in Figure \ref{fig:elmy_eda_sepos}, but not entirely independently. Figure \ref{fig:alpha_beta_lambda} shows the values of $\beta_{e}$ and $\lambda_{p_{e}}$ plotted against $\alpha_{t}$ for this dataset, including curves of best fit. Indeed, changes in $\alpha_{t}$ represent rather monotonic changes to $\beta_{e}$ and $\lambda_{p_{e}}$\footnote{Note the much stronger dependence on $\alpha_{t}$ (stronger than quadratic) for this dataset, which also includes ELMy H-modes compared to that found in the larger dataset in Section \ref{sec:thomson}, which consisted primarily of EDA H-modes. For completeness, when fit using Equation \ref{eq:lambdap_widening}, this dataset gives \{$C_{\alpha}$, $a$, $C_{\rho}$, $r$\} = \{$1.1, 2.6, 1.3 \times 10^{-2}, 0.3$\}. These values are closer to that found in \cite{eich_turbulence_2020} and \cite{faitsch_analysis_2023}, which also included both regimes, and yields a significantly more scatter-free correlation.}. When the constitutent parameters of $k_\mathrm{EM}$ and $k_\mathrm{RBM}$ are plotted against each other, the source of the break in slope when transitioning from ELMy to EDA discharges observed in Figure \ref{fig:kEM_kRBM} becomes clearer. Parameterizing the parameters plotted here by the curves of best fit with $\alpha_{t}$, i.e. $\beta_{e} = \beta_{e}(\alpha_{t})$ and $\lambda_{p_{e}} = \lambda_{p_{e}} (\alpha_{t})$, substituting into Equation \ref{eq:elmy_eda}, and solving numerically yields $\alpha_{t} = 0.53$, $\beta_{e} = 9.9 \times 10^{-5}$, and $\lambda_{p_{e}} = 2.0$ mm. This procedure yields transition values of $\alpha_{t}$ and $\beta_{e}$ close to those manually identified in \cite{faitsch_analysis_2023} and in this dataset. The existence of a critical $\lambda_{p_{e}}$ is consistent with recent work proposing an onset threshold in pressure gradient for the emergence of local ballooning modes associated with the QCE regime on AUG and JET \cite{dunne_quasi-continuous_2024}.

\section{Discussion and application to SPARC}
\label{sec:sparc_projections}

That the SepOS applies to a device like C-Mod, with a higher toroidal magnetic field, $B_{t}$, and plasma density, $n$, than AUG is important. The model gives boundaries in terms of dimensionless physics parameters, so there is no reason to expect that it should not apply to C-Mod. But that it predicts them for a much larger $n$ implies that plasma turbulence physics is dominant over atomic physics in setting operational boundaries, motivating evaluation of these dimensionless parameters for and extrapolation of the model to next-step devices, like SPARC. Interestingly, AUG, C-Mod, and SPARC form close to a geometric sequence in terms of both $B$ and $n$, with a factor of 2 -- 3, separating one from the next. For the typical AUG conditions, $B_{p}$ = 0.28 T, just under half that of typical C-Mod conditions, where $B_{p}$ = 0.57 T, which is just under three times lower than the SPARC PRD at $B_{p}$ = 1.70 T. The density minimum for the L-H transition of AUG from the SepOS model is around $0.2 \times 10^{20}$ m$^{-3}$, which is about three times less that of C-Mod at $0.6 \times 10^{20}$ m$^{-3}$, and this is slightly under three times less than that of the projected SPARC PRD at $1.5 \times 10^{20}$ m$^{-3}$. Recent work has evaluated the SepOS-predicted density minima for the L-H transition and shown that they agree with predictions from the Ryter scaling for the density minimum \cite{eich_separatrix_2024}, noting the difference in use of the line-averaged electron density in that work.

\begin{figure*}[h]
\centering
\includegraphics[width=1.8\columnwidth]{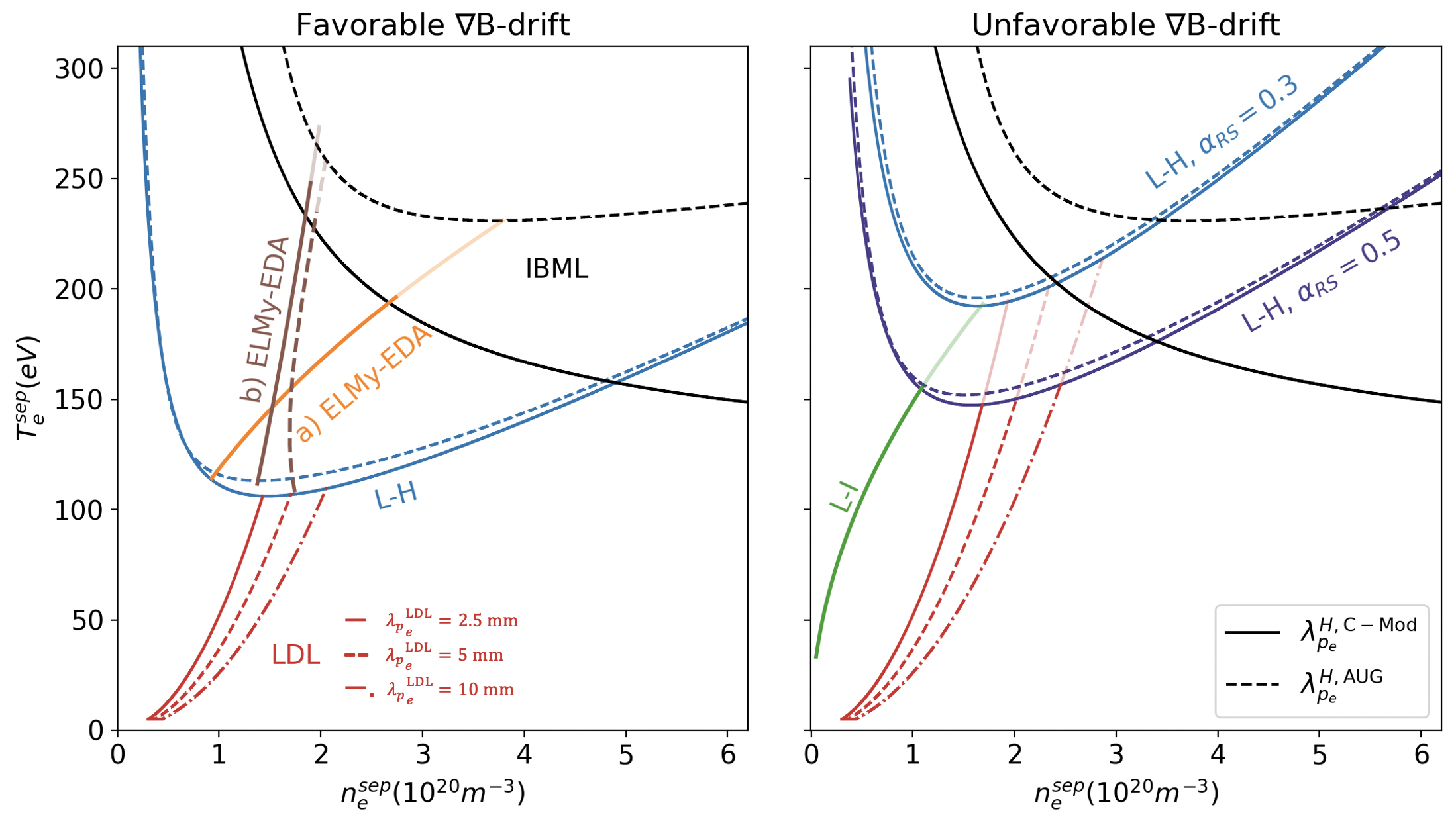}
\caption{Projected boundaries for the separatrix operational space of SPARC based on PRD parameters, using the SepOS for both the favorable $\nabla B$-drift (left) and unfavorable $\nabla B$-drift (right) directions. Both plots include the three primary SepOS boundaries. In shades of blue are the L-H transition curves. For the unfavorable $\nabla B$-drift direction, a curve is shown for $\alpha_\mathrm{RS}$ = 0.5, which is is direct extrapolation from C-Mod, and a curve is shown for $\alpha_\mathrm{RS}$ = 0.3, resulting from plugging in $\hat{q}_\mathrm{cyl} \approx 3.1$ from the PRD into the empirical expression for $\alpha_\mathrm{RS}$ from Figure \ref{fig:rf_lih}. For both plots curves are shown for the LDL (red) and the IBML (black). For the favorable $\nabla B$-drift case, the transition between the Type-I ELMy and the EDA H-mode is shown in a) orange and b) brown for different proposed transition criteria. A proposed boundary for the L-I is shown in green only for the unfavorable $\nabla B$-drift direction. Linestyles reflect different choices of $\lambda_{p_{e}}$. For boundaries dependent on H-mode scalings, including the L-H, the IBML, and criterion b) for the ELMy-EDA transition, both the scaling results from C-Mod (solid) and AUG data (dashed) are shown. For the LDL, which uses the L-mode scaling, three choices for $\lambda_{p_{e}}$ are shown, in solid, dashed, and dash-dotted with increasing $\lambda_{p_{e}}^\mathrm{LDL}$.}
\label{fig:sparc_sepos}
\end{figure*}

This validation exercise and exploration of new regime boundaries can be used to provide information about regime access and limit avoidance for next-generation devices, like SPARC \cite{creely_overview_2020}. Following on from recent work, this section shows the application of the SepOS model to the operational space of SPARC in both the favorable and unfavorable drift directions. SPARC's primary reference discharge (PRD) \cite{creely_overview_2020, Rodriguez-Fernandez_2022, body_sparc_nodate}, used to reach its mission of fusion gain, Q $>$ 2, is designed using the set of engineering parameters listed in Table \ref{tab:sparc_params}. From these engineering parameters, the DALF-normalized quantities discussed in Sections \ref{sec:forward_field_os} -- \ref{sec:elmy_eda} are calculated. As is done in these sections, combinations of these quantities are numerically solved for SPARC's parameters. These are plotted in terms of ($n_{e}^\mathrm{sep}$, $T_{e}^\mathrm{sep}$) at the separatrix for both the favorable and the unfavorable $\nabla B$-drift directions in Figure \ref{fig:sparc_sepos}. Note that the SPARC PRD is in a double null magnetic configuration while the SepOS diagrams shown here are for SPARC plasmas with the same parameters, but in either lower or upper single null. 

\begin{table}[h]
\begin{center}
\caption{Parameters for SPARC's primary reference discharge \cite{creely_overview_2020, Rodriguez-Fernandez_2022, body_sparc_nodate}}
\label{tab:sparc_params}
\begin{tabular}{cc}
Parameter & Value \\
\midrule
\midrule
$B_{t}$ (T) & 12.2 \\
\midrule
$I_{P}$ (MA) & 8.7 \\
\midrule
$B_{p}$ (T) & 1.7 \\
\midrule
R (m) & 1.85 \\
\midrule
a (m) & 0.57 \\
\midrule
$\delta$ & 0.54 \\
\midrule
$\kappa$ & 1.97 \\
\midrule
\end{tabular}
\end{center}
\end{table}

As on C-Mod and AUG, translating these boundaries into the dimensional ($n_{e}$, $T_{e}$) space for SPARC requires parametrization of the DALF-normalized quantities in terms of $n_{e}$ and $T_{e}$, as well as the variables that comprise the set of magnetic equilibrium parameters, $\mathcal{M}$. Most of these quantities have well-defined, closed-form expressions, with the exception of $\lambda_{\perp}$, as mentioned earlier. The SepOS model chooses $\lambda_{\perp} = \lambda_{p_{e}}$, but as motivated in Section \ref{sec:thomson}, there is some uncertainty about how $\lambda_{p_{e}}$ scales in the edge. As a result of this uncertainty, the SepOS boundaries are generated for different choices of $\lambda_{p_{e}}$ scalings, in both H-mode and in L-mode. In H-mode, the L-H, IBML, and criterion b) for the ELMy-EDA transition all depend on choice of $\lambda_{p_{e}}$. Figure \ref{fig:sparc_sepos} shows these curves for scalings for the H-mode $\lambda_{p_{e}}$, using the scaling coefficients from both C-Mod and AUG tabulated in the first two columns of Table \ref{tab:fit_coefficients}. For the LDL, no scaling has been generated, so a number of values for $\lambda_{p_{e}}^\mathrm{LDL}$ have been used. Their motivation is given below. The boundaries generated using threshold $\alpha_{t}$ values, i.e. the L-I transition and criterion a) for the ELMy-EDA transition, have no dependence on $\lambda_{p_{e}}$, so only one curve is generated for each. Finally, the two sets of solid and dashed curves in dark and light blue in the unfavorable $\nabla$B-drift direction account for uncertainty in the Reynolds factor, $\alpha_\mathrm{RS}$. The light blue curves take a fixed $\alpha_\mathrm{RS}$, found to best describe C-Mod data across the range of $\hat{q}_\mathrm{cyl}$, while the dark blue curves use $\alpha_\mathrm{RS}$ = 0.3, calculated from the empirical scaling found in Section \ref{sec:reverse_field_os} for $\alpha_\mathrm{RS}$ as a function of $\hat{q}_\mathrm{cyl}$. 

While both C-Mod and AUG scalings find some widening of $\lambda_{p_{e}}$ with $\alpha_{t}$ and $\rho_{s,p}$, the exponents ($a$ and $r$) are different. For data from the typical C-Mod shape shown in Figure \ref{fig:lambdap_widening}, both $a$ and $r$ are smaller than for the AUG database (although both are still positive). As for the $\alpha_{t}$ scaling, a possible explanation for the lower exponent is the lack of Type-I ELMy H-modes, which tend to have lower $\alpha_{t}$ in the typical shape. At low values of $\alpha_{t}$, the contribution from turbulence widening should be small, and without these values, it may be hard to show significant widening at higher $\alpha_{t}$. Indeed, inclusion of low $\alpha_{t}$ type-I ELMy H-modes in a scaling, as shown for the non-typical shape in Figure \ref{fig:alpha_beta_lambda} yields a much larger regression exponent. For $\rho_{s,p}$, an explanation for the low exponent could be related to measurement resolution. The smallest $\lambda_{p_{e}}$ recorded for the large typical shape database is 1.5 mm. The average channel spacing around the separatrix mapped to the midplane is somewhere between 1.2 -- 1.3 mm. As $B_{p}$ increases to $>$ 1 T, it would be hard to resolve scale lengths close to $\sim$ 1 mm, which a larger $\rho_{s,p}$ scaling might imply. To generate the LDL curves shown throughout this work, the more crude approach of simply picking $\lambda_{p_{e}}$ for non-disruptive points close to the LDL has been used. Of course, without a regression like that made for H-modes, it is impossible to make a projection for how this will scale to SPARC. As a result, three curves are generated for the LDL, corresponding to $\lambda_{p_{e}}^\mathrm{LDL} =$ 2.5, 5.0, and 10.0 mm. The first is a value extrapolated from C-Mod using a linear $B_{p}$ scaling. The third is the value used for the C-Mod SepOS curves. The second is then an intermediate value. This type of projection would benefit from a multi-machine scaling of $\lambda_{p_{e}}$ against $\alpha_{t}$ and $\rho_{s,p}$.

Figure \ref{fig:sparc_sepos} shows that the L-H boundary is not strongly sensitive to the choice of scaling for $\lambda_{p}$, especially at low and high $n_{e}$. Only at around the minimum of the curve does the C-Mod scaling predict a smaller $\lambda_{p}$ (and hence weaker $\gamma_{i}E_{ti}$), which yields a slightly lower $T_{e}$ prediction for the L-H transition. The IBML, however, shows a fairly large discrepancy between both scalings. The AUG scaling predicts a larger $\lambda_{p}$, which lowers $k_\mathrm{ideal}$, provoking the IBML at higher $T_{e}$, significantly opening up the high $n_{e}$, high $T_{e}$ operational space. The difference grows more stark as $n_{e}$ increases. While the difference in these boundaries may be large, it is not likely that operating with a separatrix density larger than $5.0 \times 10^{20}$ m$^{-3}$ in H-mode is desirable from the point of view of performance and fusion gain. 

Figure \ref{fig:sparc_sepos} also shows proposed boundaries for the ELMy-EDA transition: a) $\alpha_{t} = 0.55$ and b) $k_\mathrm{EM} = k_\mathrm{RBM}$, with both the C-Mod and AUG $\lambda_{p_{e}}$ scaling. The curves bisect the $n_{e}$, $T_{e}$ space in slightly different ways but both generally define a region at low $n_{e}$ and high $T_{e}$ where Type-I ELMs will likely be found, and a region at high $n_{e}$, low $T_{e}$ where these will likely be absent. All of these boundaries are consistent with a QCE/EDA operating point on SPARC at $n_{e}^\mathrm{sep}$ = 4.0 $\times 10^{20}$ m$^{-3}$, $T_{e}^\mathrm{sep}$ = 156 eV \cite{eich_separatrix_2024}. For discharges in the unfavorable drift direction, uncertainties in $\alpha_\mathrm{RS}$ make large differences in L-I-H access. Even changing between 50\% Reynolds stress energy transfer to 30\% represents a change in the minimum $T_{e}$ for the L-H transition, $T_{e,\mathrm{min}}$ from $\sim$150 eV to $\sim$190 eV. Interestingly though, a lower $\alpha_\mathrm{RS}$ allows easier avoidance of H-mode at higher $T_{e}$, allowing for I-mode operation at higher $n_{e}$, offering another high performance and ELM-free alternative to the type-I ELMy H-mode.

\section{Conclusions}
\label{sec:conclusions}

This work represents the first validation of the SepOS model on a device other than AUG. Using measurements from C-Mod's ETS, it provides evidence that this model for operational boundaries is applicable on a device at over double $n$ and $B$. From highly spatially resolved ETS, robust estimation of gradient scale lengths allows for high fidelity separatrix identification, a workflow very closely paralleling that used on AUG. Combining local plasma parameters and their gradient scale lengths from this approach with global engineering parameters allows for computation of control parameters extracted from equations for drift-Alfvén fluid turbulence. This allows for construction of scalings for gradient lengths, validation of proposed boundaries from the SepOS boundaries, and exploration of extension of the model to study ELM-suppressed regimes. This workflow developed on AUG and now expanded to study C-Mod data is powerful in that it allows identification of easily computable, yet still physics-based control parameters, allowing for projection on next-generation devices, like SPARC.

As was observed for H-modes on AUG and previously for L-modes on C-Mod, it is found that modifying the collisionality at the separatrix changes the local gradient scale lengths, in particular $\lambda_{p_{e}}$. Dependence is observed on $\alpha_{t}$ and $\rho_{s,p}$, which supports the body of work that both magnetic drifts and enhanced turbulence from the resistive ballooning mode work together to modify gradient scale lengths at the separatrix. The DALF-normalized quantities that underpin the SepOS model are then computed and compared. Proposed energy transfer rates are compared and found to correlate well with H-mode existence across a large range of $B_{p}$. Similarly, wavenumbers from the SepOS model are compared, and again, found to well delineate boundaries for stable L-modes and H-modes -- in particular, the L-mode density limit and the ideal ballooning MHD limit. This validation of the SepOS model on C-Mod lends confidence in using similar DALF-normalized quantities to project H-mode access and limit avoidance in next-step devices. To map these quantities onto more tangible parameters, like separatrix $n_{e}$ and $T_{e}$, which may more easily be tuned and sought by a control room operator, some work remains. The current analysis provides confidence that $\alpha_{t}$ and $\rho_{s,p}$ are useful parameters for constructing a scaling for $\lambda_{p_{e}}$ (and perhaps also $\lambda_{\perp}$), and more detailed analysis might help inform discrepancies in the scalings. Additionally, a multi-machine empirical scaling or a theory-based model for $\lambda_{\perp}$ would be pivotal for removing such uncertainties for next-step SepOS based projections. Regardless, the work also makes it clear that the SepOS model is not entirely dependent on the exact choice for the scaling, and even variation in the dependence of the scaling still permits application of the model.

As it becomes clearer that a Type-I ELMy scenario is undesirable for reactors, attention has turned to understanding access to ELM-suppressed regimes. Each ELM-suppressed regime will require its own very specific physics understanding. But, a body of evidence is growing to indicate that the physics of interest for a number of these regimes is that of the edge, and perhaps close to the separatrix. And, it may be that electromagnetic fluid drift turbulence theory applies well enough to the edge to describe some of these transitions. The DALF-normalized quantities used in this work, and in particular, $\alpha_{t}$, have proven useful in at least providing some intuition for the transition between the Type-I ELMy H-Mode and a high density, no- or small-ELM regime, like the EDA or QCE, as well as the transition to the I-mode. This intuition is built from decades of studying these regimes and operational conditions that favor them \cite{greenwald_characterization_1999, hughes_observations_2002, whyte_i-mode_2010, hubbard_edge_2011}. A key to building confidence in access to these regimes is to improve understanding of observed modes in each regime, like the quasi-coherent mode in the EDA/QCE and the weakly-coherent mode or geodesic acoustic mode in I-modes. A great deal of effort has been spent into understanding these \cite{snipes_quasi-coherent_2001, cziegler_fluctuating_2013, happel_turbulence_2016, golfinopoulos_edge_2018, manz_physical_2020, gil_stationary_2020, bielajew_edge_2022, grenfell_multi-faced_2024} and effort is beginning to connect mode properties with turbulence-related parameters from electromagnetic fluid drift turbulence theory \cite{kalis_experimental_2024}. Continuation of such efforts is crucial to development of theories for ELM avoidance and application of these models to a reactor operating scenario.

\section*{Acknowledgments}
This work was supported in part by US DOE Awards DE-SC0021629, DE-SC0014264, and DE-SC0007880 and by Commonwealth Fusion Systems. It has also has been carried out within the
framework of the EUROfusion Consortium, funded by the European Union via
the Euratom Research and Training Programme (Grant Agreement No
101052200 — EUROfusion). Views and opinions expressed are however those
of the author(s) only and do not necessarily reflect those of the
European Union or the European Commission. Neither the European Union
nor the European Commission can be held responsible for them.

\section*{References}

\bibliographystyle{unsrt}
\bibliography{references}


\end{document}